\documentclass[preprint,aps,showpacs]{revtex4}
\usepackage{epsf, epsfig, graphics,amsmath}
\begin{document}
\draft
\title{Dynamics of biomembranes with active multiple-state inclusions}
\author{Hsuan-Yi Chen$^{(1)(2)}$ and Alexander S. Mikhailov$^{(3)}$}
\affiliation{$^{(1)}$Department of Physics and Institute of
Biophysics, National Central University, Jhongli, 32001, Taiwan \\
and\\
 $^{(2)}$ Institute of Physics, Academia Sinica, Taipei, 11529,
Taiwan \\
and \\
$^{(3)}$Abteilung Physikalische Chemie, Fritz-Haber-Institut der
Max-Planck-Gesellschaft, Faradayweg 4-6, 14195 Berlin, Germany }
\date{\today}

\begin{abstract}
Nonequilibrium dynamics of biomembranes with active inclusions is
considered. The inclusions represent protein molecules which perform
cyclic internal conformational motions driven by the energy brought
with ATP ligands. As protein conformations cyclically change, this
induces hydrodynamical flows and also directly affects the local
curvature of a membrane. On the other hand, variations in the local
curvature of the membrane modify the transitions rates between
conformational states in a protein, leading to a feedback in the
considered system. Moreover, active inclusions can move diffusively
through the membrane so that surface concentration varies. The
kinetic description of this system is constructed and the stability
of the uniform stationary state is analytically investigated. We
show that, as the rate of supply of chemical energy is increased
above a certain threshold, this uniform state becomes unstable and
stationary or traveling waves spontaneously develop in the system.
Such waves are accompanied by periodic spatial variation of membrane
curvature and inclusion density. For typical parameter values, their
characteristic wavelengths are of the order of hundreds of
nanometers. For traveling waves, the characteristic frequency is of
the order of a thousand Hz or less.

\end{abstract}
\pacs{87.16.Dg, 05.40.-a, 05.70.Np}

\maketitle
 \section{Introduction}
\noindent Membranes play a fundamental role in various functions of
living cells, providing spatial compartmentalization and being
essential for signal transduction and inter- or intracellular
transport~\cite{ref:Lodish_book}. While in vivo experiments with
biomembranes are difficult because of their strong coupling to
cytoskeleton and cytoplasm, investigations with synthetic membranes
and vesicles formed by lipid bilayers can be
performed~\cite{ref:06_JP_Sackmann}. Equilibrium morphology
transitions in vesicles are well described by the elastic membrane
theory~\cite{ref:97_Adv_Phys_Seifert}. However, it becomes
increasingly clear that, under standard physiological conditions in
a living cell, biomembranes are far from the state of thermal
equilibrium.

    Lipid bilayers that constitute a biomembrane are usually including
a large number of protein molecules which may represent ion pumps,
enzymes or other molecular machines. The common property of active
protein inclusions is that they undergo cyclic conformational
changes, with each next cycle initiated by binding of an
energy-bringing ligand, typically an ATP molecule. Active
conformational motions in the inclusions are coupled to membrane
dynamics, thus bringing a membrane away from the state of
thermodynamic equilibrium.

    Theoretical investigations of biomembranes including active ion
pumps have shown~\cite{ref:96_EPL_Prost} that their fluctuational
power spectra must be very different from that of the equilibrium
membrane systems. It has been experimentally demonstrated that
active inclusions renormalize bending
rigidity~\cite{ref:01_PRE_Prost,ref:05_PRL_Prost} and surface
tension~\cite{ref:08_Prost} of biological membranes. In several
theoretical studies, it has been furthermore
predicted~\cite{ref:00_PRL_Ramaswamy,ref:04_PRL_Chen} that
instabilities of membrane shapes may be induced by the
nonequilibrium activity of the inclusions. Such instabilities could
not however so far been observed in the experiments.

    The original theoretical analysis of active membrane instabilities
has been performed~\cite{ref:00_PRL_Ramaswamy} in the framework of a
general phenomenological approach where the details of operation of
individual protein inclusions did not play a role. Effects of
conformational transitions in active protein inclusions have been
explicitly considered in the previous
publication~\cite{ref:04_PRL_Chen,ref:06_PRE_Chen} where the
equations for membrane dynamics have been complemented by a kinetic
equation for conformational transitions in molecular inclusions.
Only inclusions with two different conformational states have been
previously studied.

    However, as we point it out now, models with two-state inclusions
cannot describe generic properties of nonequilibrium membrane
systems. Indeed, when only two discrete states are present in active
element and stochastic transitions between them take place, such
transitions always satisfy the condition of detailed balance.
Therefore, for such elements, there is no principal difference
between the states of thermal equilibrium and nonequilibrium steady
states, the latter being always describable as equilibrium states
with some effective temperature.

    In order to see true differences in the behavior of equilibrium
and nonequilibrium membranes, inclusions with at least three
internal conformational states must therefore be considered. For
such inclusions, cyclic probabilities flows indicating the absence
of the detailed balance are possible. The intensity of such flows
provides direct characterization of the degree of deviation from
thermal equilibrium.

    Having this in mind, we construct here a complete kinetic description
for the membranes with multiple-state inclusions, representing a
generalization of the previous
theories~\cite{ref:96_EPL_Prost,ref:01_PRE_Prost,ref:04_PRL_Chen}.
For the general case of the $K$-state inclusions, a detailed
stability analysis is analytically performed. In addition to the
long-wavelength static instability due to the negative effective
surface tension, which has previously been known, two new
instabilities are found. They are the Turing-type static instability
with a finite spatial wavelength and the oscillatory wave-type
instability with a finite wavelength. The characteristic wavelengths
of both instabilities are of the order of hundreds of nanometers;
the characteristic frequency of the patterns emerging as a result of
the wave instability is up to a thousand Hz.

    Below in Section II the theoretical model is formulated. First the
expressions for the free energy of the system are given and then the
dynamical evolution equations are constructed. In Section III, the
slow-time limit of the model, valid when the time scales of
intramolecular processes are much shorter than the time scales of
the merging membrane patterns, is introduced and discussed. The
stability analysis of the uniform stationary state is performed in
Section IV. The results are finally discussed in Section V.

\section{The model}
\subsection{Free energy and the equilibrium state}
%
%  This part explains the Hamiltonian of the system.
%
We consider a membrane composed of lipids and active inclusions with
$K$ internal states. In this study, we assume that all inclusions
have the same orientation with respect to the membrane. The
situation where inclusions with both up and down orientations are
present can be considered in a similar way.  Before discussing the
dynamics, we first consider the free energy of this system.  At
coarse-grained level, its free energy is given by the sum of the
elastic energy of a lipid bilayer, the free energy associated with
the inclusions, and the energy of coupling between the inclusions
and the elastic membrane,
\begin{eqnarray}
F = F_{mem} + F_{inc} + F_{coup}.
\end{eqnarray}
For a quasi-flat membrane, the elastic free energy of a lipid
bilayer is
\begin{eqnarray}
  F_{mem}
= \frac{1}{2} \int d^2 r \left[ \gamma (\nabla h)^2 + \kappa (\nabla
^2 h)^2 + k_2 (\nabla ^3 h)^2 \right], \label{eq:Fmem}
\end{eqnarray}
where $h$ is the height of the membrane relative to the reference
plane, ${\bf r}=x \hat{\bf i}+y \hat {\bf j}$ is the two-dimensional
position vector in the reference plane, and $\nabla =
\partial _x \hat{\bf i} +
\partial _y \hat{\bf j}$. The coefficient $\gamma$ characterizes
surface tension of the membrane, $\kappa$ specifies the bending
rigidity of the membrane, and the coefficient $ k_2$ characterizes
its rigidity with respect to the spatial variation of the curvature.
Below, the last term is taken into account only for the
short-wavelength instabilities, where it becomes essential and
determines the instability wavelength. Since the system under
consideration is a biomimetic membrane, we take the surface tension
of typical cell membranes as the upper limit for the magnitude of
$\gamma$, i.e, $\gamma \lesssim 5 \times 10^{-3} k_BT/{\rm
nm}^2$~\cite{ref:92_BPJ_Needham}.  The coefficient $\kappa$
typically is $\kappa \sim 10 {\rm k_BT}$~\cite{ref:95_BPJ_Strey},
and $k_2 \sim a^2 \kappa \sim 300 \ {\rm k_BT \cdot nm^2}$ ($a \sim
5 {\rm nm}$ is the linear size of an inclusion).

The free energy associated with the inclusions is
\begin{eqnarray}
 F_{inc} = n_0 \int d^2 r \left\{ k_BT \left[ \sum _{\alpha} \phi _{\alpha}
 \ln \phi _{\alpha} + (1-  \sum _{\alpha} \phi _{\alpha})
 \ln (1-   \sum _{\alpha} \phi _{\alpha})\right] +  \sum _{\alpha} E_{\alpha}
 \phi _{\alpha} \right\}, \label{eq:Finc}
\end{eqnarray}
where $0\leq \phi _{\alpha} \leq 1$ is the dimensionless surface
density of inclusions in the state $\alpha$ ($\alpha =
1,2,3,...,K$), $E_{\alpha}$ is the energy of an inclusion in the
state $\alpha$, $n_0=1/a^2$. In this paper, we consider the case of
the low inclusion density and neglect therefore possible direct
lateral interactions between the inclusions.

The coupling between the inclusion density and the local membrane
curvature is
\begin{eqnarray}
 F_{coup} =  \int d^2 \ r \kappa \sum _{\alpha}c_{\alpha} \phi _{\alpha}
 \nabla ^2 h, \label{eq:Fcoup}
\end{eqnarray}
where $c_{\alpha}$ are the coefficients specifying the strength of
coupling between inclusions in different internal states $\alpha$
and the local membrane curvature. When $c_{\alpha} >0$, membrane
regions rich in the inclusions in the state $\alpha$ tend to have
$\nabla ^2 h <0$. The coefficient $c_{\alpha}$ has dimension of
inverse length, $c_{\alpha}\sim \Delta \Sigma _{\alpha}/\Sigma
_{\alpha} l_{\alpha}$ where $\Sigma _{\alpha}$ is the average
cross-sectional area, $\Delta \Sigma _{\alpha}$ is the difference
between the outer and inner leaflet surface area of a state-$\alpha$
inclusion , and $l_{\alpha}$ is the thickness (in the $z$ direction)
of the inclusion (see Appendix for the explanations). Typically,
$c_{\alpha}$ is of the order of $\leq 0.1 {\rm nm}^{-1}$.

%
%  This part discusses the equilibrium state associated with the
%  above mentioned Hamiltonian.
%
The equilibrium state of the system satisfies the conditions $\delta
F/\delta h=0$ and $\delta F/\delta \phi _{\alpha}=0$. This leads to
the equations determining the equilibrium state,
\begin{eqnarray}
 \kappa \nabla ^4 h - \gamma \nabla ^2 h -k_2 \nabla ^6 h + \kappa \sum _{\alpha}
 c_{\alpha} \nabla ^2 \phi _{\alpha} =0,
 \label{eq:dFdh1}
\end{eqnarray}
and
\begin{eqnarray}
\kappa c_{\alpha } \nabla ^2 h + k_BT n_0 \left[  \ln \phi _{\alpha}
- \ln (1-\sum _{\beta} \phi _{\beta} ) \right] + n_0 E_{\alpha} =0.
\label{eq:phi_eq}
\end{eqnarray}
Eq.~(\ref{eq:phi_eq}) gives us
\begin{eqnarray}
  \phi _{\alpha} = (1-\sum _{\beta} \phi _{\beta} ) \exp \left[
  -(E_{\alpha} + n_0^{-1} \kappa c_{\alpha} \nabla ^2 h)/k_BT \right].
\end{eqnarray}
Therefore
%$\sum _{\alpha} \phi _{\alpha} = (1-\sum _{\beta} \phi
%_{\beta} ) \sum _{\alpha} \exp \left[ -(E_{\alpha } + n_0^{-1}
%\kappa c_{\alpha} \nabla ^2 h)/k_BT \right]$.  The
the local total
density of the inclusion in thermal equilibrium is simply
\begin{eqnarray}
\phi _t = \sum _{\alpha} \phi _{\alpha} = \frac{\sum _{\alpha} \exp
\left[ -(E_{\alpha } + n_0^{-1} \kappa c_{\alpha} \nabla ^2 h)/k_BT
\right] }{1+ \sum _{\alpha} \exp \left[ -(E_{\alpha } + n_0^{-1}
\kappa c_{\alpha} \nabla ^2 h)/k_BT
 \right] }.
\end{eqnarray}
Note that the total number of inclusions present in the membrane is
$N=n_0 \int d^2r \ \phi _t$.  It depends on $E_{\alpha}$, implying
that $E_{\alpha}$ includes the contribution from the free energy
difference between an $\alpha$-state inclusion in the membrane and
an $\alpha$-state inclusion in the solvent.

Already the equilibrium state of the system may exhibit complex
structural phases such as the undulated lamellar phase, the hexatic
arrangement of inclusions, etc (see~\cite{ref:equilibrium}). These
phases are possible because of the strong coupling between the
inclusion density and the local membrane curvature, and it is has
been argued that they are important for biological
membranes~\cite{ref:04_BPJ_Turner_Sens}. Because we are primarily
interested in the nonequilibrium behavior, we limit however our
study below to the situations where the uniform flat state of the
membrane is stable under equilibrium conditions.

\subsection{Evolution equations}
If a membrane is impermeable, it can only move together with the
solvent flow. A (partially) permeable membrane can be dragged
through the solvent. Thus, the equation for the temporal evolution
of $h({\bf r},t)$ is
\begin{eqnarray}
\partial _t h = {\bf v}\cdot \hat{\bf n} -\lambda _p \left[ \frac{\delta F}{\delta h}
+ \sum _{\alpha, \ \beta} k_{\alpha \beta} n_0P_{\alpha \beta}^m
\phi _{\beta} \right], \label{eq:permeation}
\end{eqnarray}
where ${\bf v}({\bf r})$ is the local solvent flow and $\hat{\bf
n}({\bf r})$ is the unit normal vector of the membrane surface.
Because of the permeation, the solvent can ``leak'' from one side of
the membrane to the other. If the force $f$ is locally applied at a
permeable membrane, it gets dragged through the solvent with the
velocity $v =\lambda _p f$, where $\lambda _p$ is the permeation
constant. The local force $f$ is given by the expression in the
brackets in equation~(\ref{eq:permeation}). In addition to the first
contribution, directly coming from the free energy, it also includes
the second term of purely kinetic origin. When an inclusion changes
its internal conformational state, some force acting on the membrane
is generated. Such average force corresponds to the second term
inside the brackets. Here, $k_{\alpha \beta}$ is the transition rate
constant for an inclusion to go from state $\beta$ to state $\alpha$
and $P_{\alpha \beta}^m$ is the momentum that an inclusion delivers
to the membrane during a $\beta \rightarrow \alpha$ conformational
change. Typically $P_{\alpha \beta}^m$ ranges from ${\rm pN} \cdot
\mu {\rm s}$ to ${\rm pN} \cdot {\rm ms}$. In our analysis, we shall
neglect thermal noises acting on the membrane, which should be
generally present in the membrane evolution equation.

The local density of inclusions in the state $\alpha$ changes with
time because of the lateral diffusion and the conformational
transitions between the states, so that the evolution equations for
the surface densities of inclusions in the state $\alpha$ are
\begin{eqnarray}
  \partial _t \phi _{\alpha} = M_{\alpha} \nabla ^2 \frac{\delta
  F}{\delta \phi _{\alpha}} + \sum _{\beta} \left[ k_{\alpha \beta}
  \phi _{\beta} - k_{\beta \alpha} \phi _{\alpha} \right],
  \label{eq:inclusions}
\end{eqnarray}
where $M_{\alpha}$ is the mobility coefficient for the
state-$\alpha$ inclusions.

Hydrodynamic flows in the solvent are described by the modified
Stokes equation in the limit of low Reynolds number that takes into
account the elastic stress due to the membrane and the forces
resulting from the inclusion conformational changes. We have
\begin{eqnarray}
0&=&-\nabla p - \partial _z p \ \hat{\bf z} + \eta (\nabla ^2 +
\partial _z^2) {\bf v} - \delta (z-h)\frac{\delta F}{\delta h}
\hat{\bf n} \nonumber \\
&& + \sum _{\alpha \beta} k_{\alpha \beta} n_0P_{\alpha \beta} \phi
_{\beta} \left[ (1+z_{\alpha \beta}^{(u)} \nabla ^2 h) \delta
(z-z^{(u)}_{\alpha \beta}-h ) -(1-z_{\alpha \beta}^{(d)} \nabla ^2
h)\delta (z+z^{(d)}_{\alpha \beta} -h ) \right] \hat{\bf n},
\nonumber \\
 \label{eq:stokes}
\end{eqnarray}
where $\eta$ is the viscosity of the solvent. In this equation,
$\delta F/\delta h$ corresponds to the pressure jump across the
membrane due to the membrane elasticity. The last term comes from
conformational changes of the inclusions.

Because the dynamics of an inclusion in the fluid is strongly
overdamped, an inclusion exerts a zero net force on the solvent
during its conformational transition from the state $\beta$ to the
state $\alpha$. The simplest description for the effect of an
inclusion conformational change on the solvent flow is therefore
provided by a model of a force dipole. $z_{\alpha \beta} ^{(u)}$ and
$z_{\alpha \beta} ^{(d)}$ in Eq.~(\ref{eq:stokes}) are
characteristic lengths (both on the order of ${\rm nm}$) for the
force distribution during a conformational change of an inclusion
from the state $\beta$ to the state $\alpha$. As has been previously
pointed out in Ref.~\cite{ref:Lomholt}, but neglected in
Refs.~\cite{ref:05_PRL_Prost} and~\cite{ref:04_PRL_Chen}, because
force centers of the inclusions lie in the surfaces parallel to the
membrane (see Fig.~\ref{fig:dipole_density}) the local density of
the force centers in the plane $h+z_{\alpha \beta}^{(u)}$ is
$(1+z_{\alpha \beta}^{(u)} \nabla ^2 h) \delta (z-z^{(u)}_{\alpha
\beta}-h )$ and the local density of the force centers in the plane
$h-z_{\alpha \beta}^{(d)}$ is $(1-z_{\alpha \beta}^{(d)} \nabla ^2
h)\delta (z+z^{(d)}_{\alpha \beta} -h )$.

In addition, because the solvent is incompressible we have
\begin{eqnarray}
 \nabla \cdot {\bf v} + \partial _z v_z=0.
 \label{eq:incompressibility}
\end{eqnarray}

\subsection{Microscopic reversibility and detailed balance}
Microscopic reversibility imposes several constraints on $k_{\alpha
\beta}$, $P_{\alpha \beta}^m$, $P_{\alpha \beta}$, $z_{\alpha \beta}
^{(u)}$, and $z_{\alpha \beta} ^{(d)}$. The local membrane curvature
affects the energy landscape of the conformational states of an
inclusion, thus in general the transition rates $k_{\alpha \beta}$
should depend on the local membrane curvature.  To the order of
$\nabla ^2 h$ we have
\begin{eqnarray}
k_{\alpha \beta} = \Omega _{\alpha \beta}^{(0)} k_{\alpha
\beta}^{(0)} \left( 1+ \Omega _{\alpha \beta} ^{(1)} l_{\alpha
\beta} \nabla ^2 h \right). \label{eq:kalphabeta}
\end{eqnarray}
Here, the parameters $k_{\alpha \beta}^{(0)}$ and $l_{\alpha \beta}$
correspond to passive inclusions, in the absence of external energy
supply; $k_{\alpha \beta}^{(0)}$ is the equilibrium transition rate
and $l_{\alpha \beta}$ represents the characteristic length
associated with the curvature dependence of $k_{\alpha
\beta}^{(0)}$. The parameters $\Omega _{\alpha \beta}^{(0)}$ and
$\Omega _{\alpha \beta}^{(1)}$ are introduced to describe changes in
these properties when inclusions become active, i.e. when the energy
is supplied to the inclusions. Thus, for passive inclusions $\Omega
_{\alpha \beta}^{(0)}=\Omega _{\alpha \beta}^{(1)}=1$ and the
transition rates take the equilibrium form $k_{\alpha \beta} =
k_{\alpha \beta}^e = k_{\alpha \beta}^{(0)} (1+l_{\alpha \beta}
\nabla ^2 h)$.

The detailed balance condition implies (to the order $h$) the
following relationship between the equilibrium transition rates
$k_{\alpha \beta}^e$ and $k_{\beta \alpha} ^e$,
\begin{eqnarray}
 \frac{k_{\alpha \beta}^e}{k_{\beta \alpha} ^e} =
 \frac{e^{-(E_{\alpha} + n_0^{-1}\kappa c_{\alpha} \nabla ^2 h)/k_BT}}
      {e^{-(E_{\beta} + n_0^{-1}\kappa c_{\beta} \nabla ^2 h)/k_BT}}
%\nonumber \\
= e^{-(E_{\alpha} - E_{\beta})/k_BT} \left( 1-\frac{\kappa
(c_{\alpha}-c_{\beta})}{n_0 k_BT} \nabla ^2h \right).
%+ \mathcal{O}(h^2).
\end{eqnarray}
The second expression is valid as long as $\kappa (c_{\alpha} -
c_{\beta} ) \nabla ^2 h /n_0 k_BT = [\kappa (c_{\alpha} - c_{\beta}
) /n_0 k_BT]/R\ll 1$, where $R =| \nabla ^2 h|^{-1}$ is the radius
of the local membrane curvature. The above condition yields
\begin{eqnarray}
\frac{k_{\alpha \beta}^{(0)}}{k_{\beta \alpha}^{(0)}} =
e^{-(E_{\alpha}-E_{\beta})/k_BT},
\end{eqnarray}
and
\begin{eqnarray}
l_{\alpha \beta}-l_{\beta \alpha} = - \frac{\kappa
(c_{\alpha}-c_{\beta})}{n_0 k_BT}.
\end{eqnarray}

 When the system is passive, there can be no active permeation.
Therefore, we have
\begin{eqnarray}
 \sum _{\alpha \beta} k_{\alpha \beta}^e P_{\alpha \beta}^m \phi
   _{\beta}=0.
\end{eqnarray}
The requirement of the detailed balance $k_{\alpha \beta}^e \phi
_{\beta}^e = k _{\beta \alpha}^e \phi _{\alpha}^e$ yields
\begin{eqnarray}
 P_{\alpha \beta}^m = - P_{\beta \alpha}^m. \label{eq:Pm_symmetry}
\end{eqnarray}
This relationship is a direct consequence of the microscopic
reversibility of an $\alpha \rightarrow \beta$ transition: the
momentum transfer from an inclusion to the membrane during an
$\alpha \rightarrow \beta$ transition has the same magnitude but
opposite direction as the momentum transfer from an inclusion to the
membrane during a $\beta \rightarrow \alpha$ transition.

Two other conditions that must be satisfied for passive systems are
\begin{eqnarray}
 \left\{ \begin{array}{ll}
  & \sum _{\alpha \beta} k_{\alpha \beta}^e P_{\alpha \beta} (z_{\alpha \beta}^{(u)}+z_{\alpha \beta}^{(d)}) \phi
   _{\beta}=0,  \nonumber \\
\ & \nonumber \\
   & \sum _{\alpha \beta} k_{\alpha \beta}^e P_{\alpha \beta}
   ({ z_{\alpha \beta}^{(u)}}^2 - {z_{\alpha \beta}^{(d)}}^2 ) \phi
   _{\beta} = 0.
   \end{array}
   \right.
\end{eqnarray}
Again, because of the detailed balance, these two conditions lead to
\begin{eqnarray}
P_{\alpha \beta} = - P_{\beta \alpha}, \label{eq:P_symmetry}
\end{eqnarray}
and
\begin{eqnarray}
z_{\alpha \beta}^{(u)} = z_{\beta \alpha}^{(u)}, \ z_{\alpha \beta}
^{(d)} = z_{\beta \alpha}^{(d)}.
\end{eqnarray}
These conditions are also direct consequences of microscopic
reversibility.

\subsection{Final evolution equations}
By using equations~(\ref{eq:stokes})
and~(\ref{eq:incompressibility}) and applying the Fourier
transformation in the coordinate space, the flow velocity ${\bf v}$
can be expressed in terms of the hight variation $h$. Substituting
the resulting expressions into~(\ref{eq:permeation})
and~(\ref{eq:inclusions}), final evolution equations for the
membrane hight and the inclusion density in different conformation
states are obtained. They have the form
\begin{eqnarray}
\partial _t h({\bf q}, t) &=& -\lambda _p \left[
\frac{\delta F}{\delta h} + \sum _{\alpha \beta} k_{\alpha \beta}
n_0P_{\alpha \beta}^m \phi _{\beta} \right] ({\bf q},t) \nonumber
\\
&&- \frac{1}{4\eta q} \left[ \frac{\delta F}{\delta h} %
+ \sum _{\alpha \beta} k_{\alpha \beta} n_0 P_{\alpha \beta} \phi
_{\beta} (z^{(u)}_{\alpha \beta} + z^{(d)}_{\alpha \beta})q^2h  +
\sum _{\alpha \beta} k _{\alpha \beta} n_0P_{\alpha \beta} \frac{
{z_{\alpha \beta} ^{(u)}}^2 - {z_{\alpha \beta} ^{(d)}}^2 }{2}  q^2
\phi_{\beta} \right]({\bf q},t), \nonumber \\
\label{eq:h2}
\end{eqnarray}
and
\begin{eqnarray}
\partial _t \phi _{\alpha} ({\bf q},t) = - M_{\alpha} q^2
\frac{\delta F}{\delta \phi _{\alpha}} ({\bf q},t) + \sum _{\beta}
\left( k _{\alpha \beta} \phi _{\beta} - k_{\beta \alpha} \phi
_{\alpha} \right) ({\bf q},t).
\label{eq:phialpha2}
\end{eqnarray}
Here $h({\bf q},t)$ and $\phi _{\alpha}({\bf q},t)$ are the
coefficients of the Fourier expansion of the local hight $h({\bf
r},t)$ and the local densities $\phi _{\alpha}({\bf r},t)$ over the
plane waves $\exp(i{\bf q}\cdot {\bf r})$.

The contributions to the membrane dynamics given by
Eq.~(\ref{eq:h2}) come (i) from the direct momentum transfer from
the inclusions to the lipid bilayer, described by $k_{\alpha \beta}
n_0 P_{\alpha \beta}^m \phi _{\beta}$; (ii) from the action of the
active force dipoles due to the coupling between the local membrane
curvature and the force density distribution, described by
$k_{\alpha \beta} n_0 P_{\alpha \beta} \phi _{\beta} (z_{\alpha
\beta}^{(u)} + z_{\alpha \beta} ^{(d)}) q^2 h$; and (iii) from the
action of the active force quadrupoles due to the up-down asymmetry
of the shape of the inclusions, described by $k_{\alpha \beta} n_0
P_{\alpha \beta} ({z_{\alpha \beta}^{(u)}}^2 - {z_{\alpha
\beta}^{(d)}}^2)q^2 \phi _{\beta}/2$.

Evolution equations~(\ref{eq:h2}) and~(\ref{eq:phialpha2}) provide
the final concise description of the membrane dynamics coupled to
teh kinetic transitions in active multiple-state inclusions
diffusing within the membrane. The physical meanings and typical
magnitudes of the parameters entering into these equations and the
respective expressions for the free energy are summarized in
Table~I.

Equations~(\ref{eq:h2}) and (\ref{eq:phialpha2}) represent the
generalization of the equations previously obtained in the
theories~\cite{ref:00_PRL_Ramaswamy,ref:Lomholt} that neglected
conformational changes inside the inclusions, where active momentum
transfer from the inclusions to the
membrane~\cite{ref:00_PRL_Ramaswamy} and active force dipoles and
active force quadrupoles~\cite{ref:Lomholt} were both taken into
account.

\section{The limit of slow membrane dynamics}
Membrane motions with long wavelengths are slow. The characteristic
time scales of such motions are much larger than the characteristic
times corresponding to the the kinetics of internal transitions
inside the inclusions. In this situation, which we consider further
in our study, the steady nonequilibrium distribution over the
internal states of active inclusions should adiabatically following
the changes in the local membrane shape.

   For example, let us consider biomembranes with active ion pumps, such as
bacteriorhodopsin (BR) and ${\rm Ca}^{2+}$-ATPase, which have used
in the experiments~\cite{ref:01_PRE_Prost, ref:05_PRL_Prost,
ref:08_Prost}. The typical timescale for conformational transitions
inside the cycle in these ion pumps is of the order of several
milliseconds. The lateral diffusion constant for an an ion pump
inside a membrane is on the order of $1 \mu {\rm m}^2/s$. Therefore,
within a single active conformational cycle such an inclusion would
move over the distance of about several tens of nanometers inside
the membrane. Hence, if the membrane shape varies on a scale of
several hundreds of nanometers or longer, the adiabatical
approximation for the inclusion kinetics shall be applicable.  In
the steady state, local inclusion densities satisfy the equations
\begin{eqnarray}
 \sum _{\beta} (k_{\alpha \beta} \phi _{\beta} - k_{\beta \alpha }
 \phi _{\alpha}) =0 \ \ {\rm for \ all \ } \alpha.
 \label{eq:longwavelength}
\end{eqnarray}
In this paper, inclusions with cyclic transitions are considered
(Fig.2), so that $k_{\alpha \beta} = 0 $ for $\beta - \alpha \neq
\pm 1$. Therefore, Eqs.~(\ref{eq:longwavelength}) give us
\begin{eqnarray}
   k_{21} \phi _1 - k_{12} \phi _2 = k_{32} \phi _2 - k_{23} \phi _3
  = k_{43} \phi _3 - k_{34} \phi _4 = ...... %\nonumber \\
\equiv K_s(k_{\alpha \beta}) \phi _t. \label{eq:current}
\end{eqnarray}
Thus, conformational transitions are characterized by a steady
conformational current $K_s \phi _t$. When $K_s >0$ the
conformational current is along $1 \rightarrow 2 \rightarrow 3
\rightarrow ... \rightarrow K \rightarrow 1$.  The presence of such
a flow (if $K_s$ is not vanishing) indicates the absence of detailed
balance and the deviation from the state of thermal equilibrium for
the inclusions.

Slow dynamics of the system is described by two variables: membrane
height $h$ and the total density of inclusions $\phi _t=\sum
_{\alpha} \phi _{\alpha}$. From Eq.~(\ref{eq:phialpha2}), we obtain
\begin{eqnarray}
\partial _t \phi _t({\bf r},t) = \sum _{\alpha} M_{\alpha} \nabla ^2
\left( k_BT n_0 \ln \frac{\phi _{\alpha}}{1-\phi _t} + \kappa
c_{\alpha} \nabla ^2 h \right). \label{eq:phit}
\end{eqnarray}
Here $\phi _{\alpha}$ is a function of $\phi _t$ and $K_s$, we will
discuss this function later after deriving the linearized equations.
To find the equation for $h$, terms related to active forces are
simplified by introducing the active momentum transfer, the active
force dipole and the active force quadrupole associated with an
inclusion. From Eqs.~(\ref{eq:Pm_symmetry}), (\ref{eq:P_symmetry}),
and (\ref{eq:current}), we obtain
\begin{eqnarray}
\sum _{\alpha \beta} k_{\alpha \beta} P_{\alpha \beta}^m \phi
_{\beta}
= (P_{21}^m+P_{32}^m+... + P_{K \ K-1}^m+P_{1 K}^m)(k_{21}\phi _1-k_{12}\phi _2) %\nonumber \\
\equiv P_A^m K_s(k_{\alpha \beta}) \phi _t, \label{eq:apermeation}
\end{eqnarray}
\begin{eqnarray}
&&\sum _{\alpha \beta} k_{\alpha \beta} P_{\alpha \beta} (z_{\alpha
\beta}^{(u)}+z_{\alpha \beta}^{(d)} ) \phi _{\beta} \nonumber \\
&=&
\left[P_{21}(z_{21}^{(u)}+z_{21}^{(d)})+P_{32}(z_{32}^{(u)}+z_{32}^{(d)})+...
+ P_{1 K}(z_{1 K}^{(u)}+z_{1 K}^{(d)}) \right](k_{21}\phi
_1-k_{12}\phi
_2)\nonumber \\
&\equiv & p_A K_s \phi _t.  \label{eq:aforcedipole}
\end{eqnarray}
and
\begin{eqnarray}
&& \sum _{\alpha \beta} k_{\alpha \beta} P_{\alpha \beta}
\frac{{z_{\alpha \beta}^{(u)}}^2-{z_{\alpha \beta}^{(d)}}^2}{2} \phi
_{\beta} \nonumber \\
&=&
\left(P_{21}\frac{{z_{21}^{(u)}}^2-{z_{21}^{(d)}}^2}{2}+P_{32}\frac{{z_{32}^{(u)}}^2-{z_{32}^{(d)}}^2}{2}+...
+ P_{1 K}\frac{{z_{1 K}^{(u)}}^2-{z_{1
K}^{(d)}}^2}{2}\right)(k_{21}\phi _1-k_{12}\phi
_2) \nonumber \\
&\equiv & Q_A K_s \phi _t.  \label{eq:aforcequadpole}
\end{eqnarray}
Here, we have defined the active momentum transfer from an inclusion
to the membrane during one cycle $P_A^m = P_{21}^m + P_{32}^m +... +
P_{1 K}^m$, the net active force dipole of an inclusion during one
cycle $p_A = P_{21}(z_{21}^{(u)}+z_{21}^{(d)}) +
P_{32}(z_{32}^{(u)}+z_{32}^{(d)}) +... + P_{1
K}(z_{1K}^{(u)}+z_{1K}^{(d)})$, and the net active force quadrupole
of an inclusion during one cycle $Q_A =
P_{21}({z_{21}^{(u)}}^2-{z_{21}^{(d)}}^2)/2 +
P_{32}({z_{32}^{(u)}}^2-{z_{32}^{(d)}}^2)/2 +... + P_{1
K}({z_{1K}^{(u)}}^2-{z_{1K}^{(d)}}^2)/2$. Note, that the respective
terms are non-vanishing and thus contribute to the membrane dynamics
only when the detailed balance is violated and the cyclic
conformational current $K_s \phi _t$ is present. The equation of
motion for $h({\bf q},t)$ can be obtained by substituting
Eq.~(\ref{eq:apermeation}), (\ref{eq:aforcedipole}), and
Eq.~(\ref{eq:aforcequadpole}) into Eq.~(\ref{eq:h2}), yielding
\begin{eqnarray}
\partial _t h({\bf q},t) %
&=&-  \lambda _p \left[ \frac{\delta F}{\delta h} + n_0P_A^m K_s
\phi _t\right]({\bf q},t) \nonumber \\
&&- \frac{1}{4\eta q} \left[ \frac{\delta F}{\delta h} + n_0 p_A K_s
\phi _t  h q^2+ n_0 Q_A K_s \phi _t q^2 \right]({\bf q},t).
\label{eq:h}
\end{eqnarray}

Eq.~(\ref{eq:h}) indicates that the membrane acquires a nonzero
velocity due to active permeation, as one can see from $\partial _t
h = - \lambda _p n_0 P_A^m K_s \phi _t $ at $q=0$. For a permeable
membrane this drift provides clear manifestation of non-equilibrium
processes in the system under consideration.

%(** Move to discussion section!!) However, so far experiments on
%active membranes were all carried out for membranes with negligible
%permeability and this active drift could not be
%observed~\cite{ref:01_PRE_Prost}. Besides this overall motion,
%previous theoretical works have also predicted that active processes
%significantly change the fluctuation power spectrum of the
%membrane~\cite{ref:00_PRL_Ramaswamy,ref:05_PRL_Prost,ref:08_Prost},
%and the possibility of having activity-induced instabilities for a
%flat membrane~\cite{ref:00_PRL_Ramaswamy,ref:04_PRL_Chen}.

\section{Nonequilibrium instabilities}
To analyze possible instabilities of the flat membrane, we should
linearize the equations of motion by expanding $h({\bf r},t)$ and
$\phi _t({\bf r},t)$ around their uniform solutions $h_0(t)$ and
$\phi _{t0}$,
\begin{eqnarray}
 &h({\bf r},t)& = h_0(t) + \delta h({\bf r},t), \nonumber \\
 &\phi _t({\bf r},t)& = \phi _{t0} + \delta \phi _t({\bf r},t).
\end{eqnarray}
Note that $\delta h({\bf q},t) = h({\bf q},t)$ and $\delta \phi
_t({\bf q},t) = \phi _t ({\bf q},t)$ for $q\neq 0$.

First, we consider the linear elasticity of the system. One can
express $\delta F/\delta h$ as
\begin{eqnarray}
\frac{\delta F}{\delta h({\bf r})} %
&=& (\kappa \nabla^4 - \gamma \nabla ^2 - k_2 \nabla ^6) h + \sum
_{\alpha} \kappa c_{\alpha} \nabla ^2 \left( \frac{\phi
_{\alpha}}{\phi _t} \phi _t \right). \nonumber
\end{eqnarray}
At small $\nabla ^2 h$, we can further express $\phi _{\alpha} /\phi
_t$ as
\begin{eqnarray}
\frac{\phi _{\alpha}}{\phi _t} = \psi _{\alpha}\left( 1 + \l_{\psi
\alpha} \nabla ^2 h + ... \right), \label{eq:psi}
\end{eqnarray}
where $\psi _{\alpha}$ is dimensionless and $l_{\psi \alpha}$ is a
length that characterizes the curvature dependence of $\phi
_{\alpha}/\phi _t$.  Because all coordinate dependence of $\phi
_{\alpha}/\phi _t$ comes from the local curvature dependence of the
transition rates, both $\psi _{\alpha}$ and $l_{\psi \alpha}$ are
independent of ${\bf r}$, but they depend on $\Omega _{\alpha \beta}
^{(0)}$ and $\Omega _{\alpha \beta} ^{(1)}$, i.e., on the strength
of the external energy drive. From Eq.~(\ref{eq:psi}), $\nabla ^2
(\phi _t \phi _{\alpha}/ \phi _t)= \psi _{\alpha} \nabla ^2 \phi _t
+ \psi _{\alpha } l _{\psi \alpha} \phi _{t0} \nabla ^4 h+ ...$.  As
a result
\begin{eqnarray}
\frac{\delta F}{\delta h({\bf r})} %
&=& \kappa (1 + \sum _{\alpha} c_{\alpha} \psi _{\alpha} l_{\psi
\alpha} \phi _{t0}) \nabla ^4 h - \gamma \nabla ^2 h - k_2 \nabla ^6
h + \sum _{\alpha} \kappa c_{\alpha} \psi _{\alpha} \nabla ^2 \phi
_t \nonumber \\
&=& \kappa _{eff} \nabla ^4 h - \gamma \nabla ^4 h - k_2 \nabla ^6 h
+ \kappa _{eff} c_{eff}^h \nabla ^2 \phi _t. \label{eq:dFdhexpand}
\end{eqnarray}
Here we have introduced the effective membrane bending rigidity
$\kappa _{eff} =\kappa (1+ \sum _{\alpha} c_{\alpha}\psi _{\alpha}
l_{\psi \alpha} \phi _{t0})$ and the effective inclusion-membrane
elastic coupling constant $c_{eff}^h =\sum_{\alpha} c_{\alpha}
\psi_{\alpha} \kappa /\kappa _{eff}$.  Note that, besides
$c_{\alpha}$, $\kappa _{eff}$ and $c_{eff}^h$ also depend on $\Omega
_{\alpha \beta}^{(0)}$ and $\Omega _{\alpha \beta}^{(1)}$ through
$\psi _{\alpha}$ and $l_{\psi \alpha}$.  These effective elastic
moduli are different from their equilibrium values because the
relative inclusion population $\phi _{\alpha} / \phi _t$ is
different from the equilibrium distribution.

Next we consider the equation for $\delta \phi _t$.  From
Eq.~(\ref{eq:psi}),
\begin{eqnarray}
 \nabla ^2 \left( \ln \frac{\phi _{\alpha}}{1-\phi _t} \right)
 &=& \nabla ^2 \left( \ln \frac{\phi _{\alpha}}{\phi _t} + \ln \frac{\phi _t}{1-
 \phi_t} \right) \nonumber \\
% &=& \nabla ^2 \left\{ \ln \left[ \psi_{\alpha}(1+ l_{\psi \alpha} \nabla ^2 h)
% \right] + \ln \left( \frac{\phi _{t0} + \delta \phi _t}{1-\phi _{t0} - \delta \phi
% _t}\right) \right\} \nonumber \\
 &=& l_{\psi \alpha} \nabla ^4 h +
 \left( \frac{1}{\phi _{t0}} + \frac{1}{1-\phi _{t0}}\right) \nabla ^2 \delta \phi _t
 \nonumber
\end{eqnarray}
Substituting the above expression into Eq.~(\ref{eq:phit}), one
finds
\begin{eqnarray}
  \partial _t \delta \phi _t &=& \sum _{\alpha} M_{\alpha} \left\{
  n_0 k_BT \left[l_{\psi \alpha} \nabla ^4 h
  + \left( \frac{1}{\phi _{t0}}
  + \frac{1}{1-\phi _{t0}} \right)\nabla ^2 \delta \phi _t \right]
  + \kappa c_{\alpha} \nabla ^4 h \right\}.
  \nonumber
\end{eqnarray}
It is convenient to introduce the effective mobility
%\begin{eqnarray}
$M_{eff}\equiv \sum _{\alpha} M_{\alpha}$,
%\end{eqnarray}
the effective susceptibility of the inclusions
%\begin{eqnarray}
$\chi_{eff} \equiv n_0 k_BT \left[ 1/ \phi _{t0} + 1/(1 - \phi
_{t0}) \right]$
%\end{eqnarray}
and the effective inclusion-membrane elastic coupling constant
$c_{eff}^{\phi}$ associated with inclusion diffusion,
%\begin{eqnarray}
$M_{eff} \kappa _{eff} c_{eff}^{\phi} \equiv \sum _{\alpha}
M_{\alpha} \left( \kappa c_{\alpha } + n_0 k_BT l_{\psi \alpha}
\right)$.
%\end{eqnarray}
Similar to $\kappa _{eff}$ and $c_{eff}^h$, $c_{eff}^{\phi}$ depends
on $\Omega _{\alpha \beta}^{(0)}$ and $\Omega _{\alpha
\beta}^{(1)}$. Now the equation for $\partial _t \phi _t({\bf q},t)$
can be written as
\begin{eqnarray}
\partial _t \phi _t ({\bf q},t)=  -M_{eff} \left(
\chi _{eff} q^2 \phi _t ({\bf q},t) - \kappa _{eff} c_{eff}^{\phi}
q^4 h({\bf q},t) \right). \label{eq:phit3}
\end{eqnarray}

To obtain the linearized equation for $\partial h({\bf
q},t)/\partial t$, we expand the conformational current to the order
of $h$ as $K_s(k_{\alpha \beta}) = K_0(1+l_ K \nabla ^2 h+ ...)$ and
obtain
\begin{eqnarray}
K_s(k_{\alpha \beta}) \phi _t %
&=& K_0 \phi _{t0} + K_0 \delta \phi _t + K_0 l_K \phi _{t0}
\nabla^2 h, \label{eq:Csphitexpand}
\end{eqnarray}
where $K_0$ and $l_K$ are both independent of ${\bf r}$, but they
depend on $\Omega _{\alpha \beta }^{(0)}$ and $\Omega _{\alpha \beta
}^{(1)}$. Substituting Eq.~(\ref{eq:dFdhexpand}) and
(\ref{eq:Csphitexpand}) into Eq.~(\ref{eq:h}), the linearized
equation of motion for $h({\bf q},t)$ in the momentum space is
found,
\begin{eqnarray}
\partial _t h({\bf q},t)
&=&- \lambda _p \left[ \left(  \gamma _a ^p q^2 + \kappa _{eff} q^4
+ k_2 q^6 \right) h({\bf q},t) -
\left( \kappa _{eff} c_{eff}^h q^2 - n_0 P_A^m K_0 \right) \phi _t ({\bf q},t) \right\} \nonumber \\
&&- \frac{1}{4 \eta q} \left[ \left( \gamma _a q^2 + \kappa _a q^4 +
k_2 q^6 \right) h({\bf q},t) - \kappa _{eff} c_a  q^2 \phi _t ({\bf
q},t) \right]. \label{eq:h3}
\end{eqnarray}
In this equation, following notations have been introduced:
\begin{eqnarray}
\gamma _a^p = \gamma - n_0 P_A^m K_0l_K \phi_{t0}
\end{eqnarray}
is the surface tension renormalized by active forces that inclusions
act on the membrane,
\begin{eqnarray}
\gamma _a = \gamma + n_0 p_A K_0 \phi _{t0} \label{eq:gammaa}
\end{eqnarray}
is the surface tension renormalized by active force dipoles,
\begin{eqnarray}
\kappa _a = \kappa _{eff} - n_0 Q_A K_0l_K \phi _{t0}
\label{eq:kappaa}
\end{eqnarray}
is the bending rigidity renormalized by curvature-induced active
quadrupoles, and
\begin{eqnarray}
\kappa _{eff} c_a = \kappa _{eff} c_{eff}^h - n_0Q_AK_0
\label{eq:ca}
\end{eqnarray}
defines the renormalized inclusion-curvature coupling due to
curvature-independent active force quadrupoles.

  Since experiments are usually carried out
for the membranes with negligible
permeability~\cite{ref:01_PRE_Prost, ref:05_PRL_Prost,
ref:08_Prost}, we consider below only instabilities of impermeable
membranes with $\lambda _p =0$.

It is convenient to write the derived the equations in the matrix
form
\begin{eqnarray}
\partial _t \left( \begin{array}{l}
 h({\bf q},t) \\
 \phi _t ({\bf q},t)
 \end{array}
 \right) = \left( \begin{array}{ll}
 M_{hh}(q) & M_{h\phi}(q) \\
 M_{\phi h}(q) & M_{\phi \phi}(q)
 \end{array}
 \right)
 \left( \begin{array}{l}
 h({\bf q},t) \\
 \phi _t ({\bf q},t)
 \end{array}
 \right),
 \label{eq:matrix}
\end{eqnarray}
where
\begin{eqnarray}
M_{hh} (q) &=& - \frac{q}{4\eta } (\gamma _a + \kappa _a q^2 + k_2 q^4),\nonumber \\
M_{h \phi}(q) &=& \frac{q}{4 \eta} \kappa _{eff} c_a , \nonumber \\
M_{\phi h} (q)&=& M_{eff} \kappa _{eff} c_{eff}^{\phi} q^4 , \nonumber \\
M_{\phi \phi } (q) &=& - M_{eff} \chi _{eff} q^2.
\label{eq:matrix_elements}
\end{eqnarray}
The definitions and physical meanings of the parameters in
Eq.~(\ref{eq:matrix})~(\ref{eq:matrix_elements}) are summarized in
Table~II.

To analyze the linear stability of the system, we seek solutions of
Eq.~(\ref{eq:matrix}) of the form $h({\bf q},t), \ \phi _t ({\bf
q},t) \sim e^{\lambda t}$. The characteristic equation for $\lambda$
is
\begin{eqnarray}
\lambda ^2 - (M_{hh}+M_{\phi \phi}) \lambda + (M_{hh}M_{\phi \phi} -
M_{\phi h} M_{h \phi}) =0. \label{eq:quadratic}
\end{eqnarray}
Let the two solutions of the above equation be $\lambda _1 ({\bf
q})$, $\lambda _2({\bf q})$,  then
\begin{eqnarray}
%\left\{ \begin{array}{l}
         \lambda _1 + \lambda _2 &=& M_{hh}+M_{\phi \phi} = - \frac{1}{4\eta }
 \left(\gamma _a q  + 4 \eta M_{eff}\chi _{eff} q^2 + \kappa _a q^3 + k_2
 q^5\right),  \nonumber \\
         \lambda _1 \lambda _2 &=& M_{hh} M_{\phi \phi} - M_{h \phi}
         M_{\phi h} = \frac{M_{eff} \chi_{efff}}{4\eta}
 \left( \gamma _a q^3  + \tilde{\kappa} _a q^5 + k_2 q^7 \right),
%        \end{array}  \right.
 \label{eq:lambdas}
\end{eqnarray}
where
\begin{eqnarray}
\tilde{\kappa} _a = \kappa _a - \frac{\kappa _{eff}^2 c_a
c_{eff}^{\phi} }{\chi _{eff}}. \label{eq:tildekappaa}
\end{eqnarray}
The following instabilities can occur in this system: \\
(i) A long-wavelength instability at $q=0$ when one of the $\lambda$
becomes zero at $q=0$ and both $\lambda _1$, $\lambda _2$ are
negative for
any nonzero $q$.\\
(ii) An oscillatory wave instability (a Hopf bifurcation with a
finite wavenumber) at $q=q_h$, when $\lambda _1 + \lambda _2 =
M_{hh}(q) + M_{\phi \phi}(q)<0$ for all $q$ except $q=q_h$ and
$\lambda _1 \lambda _2 = M_{hh}(q) M_{\phi \phi}(q) - M_{h \phi}(q)
M_{\phi h}(q)>0$ for all $q$.  The frequency of the wave at the
onset of
the instability is $\omega /2 \pi = \sqrt{\lambda _1 (q_h)\lambda _2 (q_h)}/2 \pi$.\\
(iii) A static Turing-type instability when $\lambda _1 + \lambda
_2<0$
for all $q$ and $\lambda _1 \lambda _2 >0$ for all $q$ except $q=q_s$. \\

The typical magnitudes of the parameters that are important for
stability analysis are summarized in Table~III.
Figure~\ref{fig:phase} shows the phase diagrams for systems with
typical parameters.  Varying the parameters does not change the
qualitative features of the phase diagrams.  The ordinates in
Fig.~\ref{fig:phase} represent the strength of curvature-dependent
pumping, and the abscissas represent the strength of
curvature-independent pumping. The state of thermal equilibrium
corresponds to the origin of coordinates; as the system is driven
out of equilibrium, the state of the system moves along a straight
line with the slope $l_K/a$. Generally, coefficients $\kappa
_{eff}$, $c_{eff}^{\phi}$, and $c_{eff}^{h}$ also vary as the system
is being moved away of equilibrium; for simplicity we take however
constant values $\kappa _{eff} = 5 k_BT$, and
$c_{eff}^{\phi}=c_{eff}^h=0.1{\rm nm}^{-1}$ when the phase
boundaries below in Fig.~\ref{fig:phase} are determined.

\subsection{The long-wavelength static instability}
%(**Move to discussion section) It has been reported
%in~\cite{ref:08_Prost} that the fluctuation power spectrum of fluid
%vesicles containing active BRs indicates that BR activities reduce
%surface tension of the vesicle. This observation also indicates the
%possibility of negative surface tension induced by high density
%active BRs. In this situation, the membrane should be unstable
%against long wavelength fluctuations.

According to Eq.~(\ref{eq:lambdas}), this instability occurs when
$\gamma _a = \gamma + n_0 p_A K_0 \phi _{t0}<0$, i.e. when the
effective surface tension becomes negative. As shown in
Fig.~\ref{fig:phase}, for systems with $p_A<0$ this instability is
possible at sufficiently large $K_0$. Using as typical values $|p_A|
\sim {\rm nm \cdot pN} \cdot \mu {\rm s}$ to ${\rm nm \cdot pN \cdot
ms}$, $K_0 \lesssim {\rm ms}^{-1}$ and $n_0 = a ^{-2} \sim 1/25 {\rm
nm}^2$, we find that to get $\gamma _a <0$ for relatively low
inclusion density, such as $\phi _{t0} \lesssim 10^{-1}$, a vesicle
with the bare surface tension $\gamma \lesssim 10^{-3} k_BT/{\rm
nm}^2$ is needed. In the experiments~\cite{ref:08_Prost}, the
measured surface tension of a vesicle with passive BRs was $\gamma
\sim 10^{-4}k_BT/{\rm nm}^2$. Thus, it should be possible to observe
this long-wavelength instability in the experiments.

\subsection{The finite-wavelength instabilities}
Besides the long-wavelength instability corresponding to negative
$\gamma _a$, there are also instabilities when $\kappa _a$ or
$\tilde{\kappa}_a$ become negative. If $\gamma _a$ is positive,
these instabilities correspond to the growth of modes with finite
wavelengths, because long-wavelength fluctuations are suppressed by
the membrane tension.

(i) When
\begin{eqnarray}
\tilde{\kappa} _a = -2\sqrt{\gamma _a k_2}, \label{eq:tildekappaac}
\end{eqnarray}
the static {\em Turing-type instability} takes place.  Its
characteristic wavenumber is
\begin{eqnarray}
q_s=(\gamma _a/k_2)^{1/4}.
\end{eqnarray}
As follows from Eqs.~(\ref{eq:gammaa}),~(\ref{eq:kappaa})
and~(\ref{eq:tildekappaa}), Eq.~(\ref{eq:tildekappaac}) is
equivalent to
\begin{eqnarray}
&&  n_0 Q_A K_0 \left( l_K \phi _{t0} - \frac{\kappa
_{eff}c_{eff}^{\phi}}{\chi _{eff}} \right) - 2 \sqrt{ (\gamma + n_0
p_AK_0 \phi _{t0})k_2 } = \kappa _{eff} - \frac{\kappa _{eff}^2
c_{eff}^{\phi}c_{eff}^h}{\chi _{eff}}.
\label{eq:Turing}
\end{eqnarray}
As seen in Fig.~{\ref{fig:static_inst}, two conditions are essential
for the occurence of this instability.  The first is that the
difference $\kappa _{eff} - \kappa _{eff}^2 c_{ef}^{\phi} c_{eff}^h
/\chi _{eff}$ should be small, implying that the curved membrane
regions are strongly attracting the inclusions. The second property
is that the combination $n_0 Q_A K_0 (l_K \phi _{t0} - \kappa _{eff}
c_{eff}^{\phi}/\chi _{eff} )$ should be large and positive, implying
strong curvature-induced active force quadrupoles. In the phase
diagram in  Fig.~\ref{fig:phase}, the Turing-type instability is
located where $K_0l_K$ is large.

The characteristic wavelength at the onset of the Turing-type
instability can be estimated. Taking $\gamma _a \sim \gamma \lesssim
5 \times 10^{-3} k_BT/{\rm nm}^2$, $k_2 \sim 300 k_BT \cdot {\rm
nm}^2$, we find that this finite-wavelength instability has a
characteristic wavelength of $2 \pi /q_s = 2 \pi /(\gamma _a
/k_2)^{1/4} \gtrsim 10 {\rm nm} - 10^2 {\rm nm}$.  Numerically
computed wavelengths for this instability are shown in
Fig.~\ref{fig:length}.  Note also that, according to
Fig.~\ref{fig:phase} and Fig.~\ref{fig:length}, for $p_A<0$ only the
membranes with inclusions that have large $l_0$ (i.e., the activity
of these inclusions is induced by membrane curvature) may exhibit
the Turing-type instability, otherwise long-wavelength instability
is taking place.

(ii) When $\kappa _a = \kappa _a ^c$, the oscillatory {\em wave
instability} with the characteristic wavevector $q_h$ takes place.
The values of $\kappa _a$ and $q_h$ are given by the solutions of
two equations,
\begin{eqnarray}
q_h = \sqrt{\frac{- \kappa _a^c}{6k_2} + \sqrt{\left(\frac{-\kappa
_a^c}{6k_2}\right)^2 + \frac{\gamma _a}{3k_2}} }, \label{eq:qh}
\end{eqnarray}
and
\begin{eqnarray}
2q_h ^3 +  \frac{\kappa _a^c}{k_2}q_h + \frac{2\eta M_{eff} \chi
_{eff}}{k_2}=0. \label{eq:cond_hopf_np}
\end{eqnarray}
 Because $q_h>0$, Eq.~(\ref{eq:qh})
requires $\kappa _a ^c$ to be negative. Generally, the above
equations for $\kappa _a^c$ and $q_h$ have to be solved numerically.
Fig.~\ref{fig:phase} shows that this instability may occur only when
$p_A$, $Q_A$, and $c_{eff}^h$, $c_{eff}^{\phi}$ all have the same
sign, and the curvature-induced pumping is not too strong (i.e.,
$l_K$ is not very large).  This condition suggests the following
picture for the mechanism of wave instability.  As shown in
Fig.~\ref{fig:wave_inst}, first, inclusions are attracted to the
regions with their preferred membrane curvature; but the active
forces from the inclusions in these regions produce curvature that
the inclusion dislike (thus $l_K$ cannot be large). As a result, the
inclusions move back and forth, chasing regions with preferred
membrane curvature. This leads to an oscillatory instability with
the characteristic wavelength which is large as compared to the
molecular scale.

The characteristic wavelength at the onset of wave instability is
about $2 \pi / q_h \sim 100 {\rm nm}$. Substituting $q_h$ to $\omega
_h = \sqrt{\lambda _1 (q_h) \lambda _2 (q_h)}$, the characteristic
time at the onset of wave instability is estimated to be of the
order of a millisecond (see Fig.~\ref{fig:time} for numerically
computed characteristic time at the onset of wave instability). This
is already on the border of the validity of the approximation of
slow membrane dynamics, because such characteristic time is
comparable with the time scale of conformational transitions in
individual active inclusions. Therefore, our analysis of the wave
instability should be viewed as only providing an indication that
such an instability may exist. The accurate quantitative analysis of
this instability should be performed without assuming the slowness
of membrane dynamics; it will be undertaken in a separate
publication.~\cite{future}

\section{Discussion}
  We have constructed a complete kinetic description for membranes
with multiple-state active inclusions.  This work generalizes
previous theories on active
membranes~\cite{ref:00_PRL_Ramaswamy,ref:04_PRL_Chen,ref:Lomholt}.
Although we assume that all inclusions have the same orientation
with respect to the membrane, the situation where inclusions with
both orientations are present is a simple extension of this work.
The active forcing resulting from the conformational changes of the
inclusions affecting membrane dynamics through a direct force acting
on the membrane (active permeation), a force dipole due to density
asymmetry induced by membrane curvature, and a force quadrupole, are
all present in  Eq.~(\ref{eq:h2}).  Comparing to previous
theoretical works on active membranes, Eq.~(\ref{eq:h2}) directly
connects the active forcing with each conformational change of the
inclusions.

In the limit of slow membrane dynamics, the distribution over the
internal states of active inclusions adiabatically following the
changes in the local membrane shape. Our theory recovers the active
force dipoles and active force quadrupoles introduced in previous
theories~\cite{ref:01_PRE_Prost,ref:Lomholt}. Furthermore, the
linearized equations of
motion~(\ref{eq:matrix})(\ref{eq:matrix_elements}) reveal that the
effective elastic moduli of the membrane are renormalized both by
the active forcing resulting from inclusion conformation changes and
the distribution over the internal states of the inclusions.  Thus,
our linearized equations of motion can be applied to analyze the
experimentally observed renormalization of the membrane
rigidity~\cite{ref:05_PRL_Prost} and surface
tension~\cite{ref:08_Prost} by inclusion activities.  It would be
interesting to see how the pumping effect due to the nonequilibrium
forcing and the non-pumping effect due to the nonequilibrium
inclusion distribution over the internal states contribute to the
observed mechanical properties of an active membrane.~\cite{future}

We also show that when the system is driven sufficiently far away
from the equilibrium state, a uniform membrane becomes unstable and,
besides a long-wavelength instability arising from negative
effective surface tension, static Turing-type or traveling waves
spontaneously develop in the system. Figure~\ref{fig:phase} shows
that the possible instabilities for an active membrane depend on the
signs of the active force dipole $p_A$, active force quadrupole
$Q_A$, and whether inclusion activity is sensitive to the local
membrane curvature (described by the magnitude of the parameter
$l_K$). For inclusions whose activity is highly sensitive to the
local membrane curvature, it is possible for a static Turing-type
instability with a characteristic length of the order of hundreds
nanometers to take place.  For inclusions whose $p_A$, $Q_A$,
renormalized elastic constants $c_{eff}^h$ and $c_{eff}^{\phi}$ all
having the same sign and activity is not very sensitive to the local
membrane curvature, it is possible for a wave-instability to take
place. The characteristic length of the traveling wave is of the
order of hundreds of nanometers and characteristic frequency is of
the order of a thousand Hz or less.

For traveling waves with frequency of the order of a thousand Hz,
the characteristic time is already comparable with the time scale of
conformational transitions in individual active inclusions.
Therefore, in the accurate quantitative analysis of this type of
waves, the inclusion distribution over the internal states should
not adiabatically follow the changes in the local membrane shape.
Such analysis should start from Eq.~(\ref{eq:h2}), waves with high
frequency and wavelength of the order of hundreds of nanometers
should manifest themselves as the synchronization of inclusion
conformational cycles.~\cite{future}

We believe our theory will inspire new experiments on the active
membranes. Once $p_A$ and $Q_A$ of the inclusions are identified
through the measurements of the surface tension and the bending
modulus of the passive and active membranes, the criterions for the
instabilities can be deduced from our theory. The static Turing
instability may be observed by optical microscope, and both
stationary and traveling waves can be detected by dynamics
scattering experiments.  In the future, it is also possible to
design ``active vesicles'' whose morphology and mobility are
controlled by the activities of the inclusions.  Such active
vesicles may harvest the energy supply in the environment, turn
chemical energy into mechanical energy, and act as membrane
machines.

\section*{Acknowledgement}
HYC thanks stimulating discussions with J-F. Joanny and P.
Bassereau.  HYC is supported by National Science Council of the
Republic of China (Taiwan) under grant number NSC 96-2628-M-008 -001
-MY2.

\section*{Appendix}
In this appendix we give a brief calculation that relates the shape
of a conical state-$\alpha$ inclusion to its coupling constant
$c_{\alpha}$.

As shown in figure~\ref{fig:c_alpha}, the outer radius, inner
radius, and the thickness of a state-$\alpha$ inclusion are
$r^{out}_{\alpha}$, $r^{in}_{\alpha}$, and $l_{\alpha}$,
respectively.  The preferred radius of curvature $R_{\alpha}$ is
related to these quantities by
\begin{eqnarray}
\frac{r^{out}_{\alpha} }{R_{\alpha}+l_{\alpha}
/2}=\frac{r^{in}_{\alpha}}{R_{\alpha}-l_{\alpha}/2}. \nonumber
\end{eqnarray}
That is,
\begin{eqnarray}
R_{\alpha}=
\frac{l_{\alpha}(r^{out}_{\alpha}+r^{in}_{\alpha})/2}{r^{out}_{\alpha}-r^{in}_{\alpha}}
=\frac{(r^{out}_{\alpha}+r^{in}_{\alpha})^2l_{\alpha}/2}{({r^{out}_{\alpha}}^2-{r^{in}_{\alpha}}^2)}
\sim \frac{\Sigma _{\alpha} l_{\alpha}}{\Delta \Sigma _{\alpha}},
\end{eqnarray}
where $\Sigma _{\alpha}$ is the average cross-sectional area of a
state-$\alpha$ inclusion, $\Delta \Sigma _{\alpha}$ is the
difference of cross-sectional area between the outer and inner
surfaces of a state-$\alpha$ inclusion. The coupling constant
$c_{\alpha}$ is simply $c_{\alpha} \sim 1/R_{\alpha}$. From
Eq.~(\ref{eq:Fmem})(\ref{eq:Fcoup}), regions rich in $c_{\alpha}
>0$ inclusions tend to have $\nabla ^2 h <0$, this is illustrated in
Fig.~\ref{fig:e_couplings}.

\newpage

%
% Here are the tables
%

%
%Table 1: notations in Eq.~(\ref{eq:Fmem})(\ref{eq:Finc})(\ref{eq:Fcoup})(\ref{eq:h2})(\ref{eq:phialpha2})\\
%

\begin{table}[h]
\caption{notations in
Eq.~(\ref{eq:Fmem})(\ref{eq:Finc})(\ref{eq:Fcoup})(\ref{eq:h2})(\ref{eq:phialpha2})
\label{table:eq13}}
\begin{ruledtabular}
\begin{tabular}{c|l|l}
 symbol \  & physical meaning & typical value \\ \hline
 $\kappa$ & membrane bending modulus & $\sim 10 k_BT$\\
 $\gamma$ & membrane surface tension & $\lesssim 5 \times 10^{-3} k_BT / {\rm  nm}^2$ \\
 $c_{\alpha}$ & inclusion-curvature coupling constant & $\lesssim 0.1 {\rm nm}^{-1}$ \\
 $k_2$ & higher bending modulus for the membrane & $\sim 300 k_BT \cdot {\rm nm}^2$\\
 $a$ & lateral linear size of an inclusion & $ {\rm 5 nm}$ \\
 $n_0$ &  $ 1/a^2$ & $\sim {\rm 0.04 nm^{-2}}$ \\
 $\eta $& solvent viscosity & $\sim 0.01-0.1 {\rm g/cm\cdot s}$\\
 $k_{\alpha \beta}$& inclusion conformational transition rate & $\sim {\rm ms}^{-1} - {\rm \mu s}^{-1}$\\
 $P^m_{\alpha \beta}$ & momentum transfer to the membrane for $\beta \rightarrow \alpha$ inclusion &
                                                        $ \sim {\rm pN \cdot \mu s} - {\rm pN \cdot ms}$ \\
                    \ & conformational change & \ \\
 $P_{\alpha \beta}$ & momentum transfer to the solvent for $\beta \rightarrow \alpha$ inclusion &
                                                        $ \sim {\rm pN \cdot \mu s} - {\rm pN \cdot ms}$ \\
                    \ & conformational change & \ \\
 $z_{\alpha \beta}^{(u)}, \ z_{\alpha \beta}^{(d)}$ & characteristic lengths associated with
                                                                                & $\sim {\rm nm}$ \\
                          \ & active force during $\beta \rightarrow \alpha$ transition & \ \\
\end{tabular}
\end{ruledtabular}
\end{table}

\newpage

%
%Table 2: notations in Eq.~(\ref{eq:matrix})\\
%

\begin{table}[h]
\caption{notations in Eq.~(\ref{eq:matrix}) \label{table:matrix}}
\begin{ruledtabular}
\begin{tabular}{c|l|l}
symbols \  &  definition & physical meaning \\ \hline %
$\psi_{\alpha}$, $l_{\psi \alpha}$
                 & $\phi _{\alpha} / \phi _t = \psi _{\alpha} (1+ l_{\psi \alpha } \nabla ^2 h)$ %
                                        & represent the relative density of state-$\alpha$ inclusions \\

$\kappa _{eff}$  & $\kappa (1+ \sum _{\alpha} \psi _{\alpha} l_{\psi \alpha} \phi _{t0})$ %
                                        & effective bending rigidity in $\delta F/\delta h$\\
$c_{eff}^h$      & $\sum _{\alpha} c_{\alpha} \psi _{\alpha} \kappa / \kappa _{eff}$ %
                                        & inclusion-curvature coupling in $\delta F/\delta h$ \\
$c_{eff}^{\phi}$ & $\sum _{\alpha} M_{\alpha} (\kappa c_{\alpha} + n_0k_BTl_{\phi \alpha})/M_{eff}\kappa_{eff}$ %
                                        & inclusion-curvature coupling in $\partial \phi _t /\partial t$\\
$K_0$            & $K_s = K_0 (1+ l_K \nabla ^2h)$
                                        & $K_0 \phi _t$ is the curvature-independent conformational \\
                 &                      & current\\
$l_K$            & $K_s = K_0 (1+ l_K \nabla ^2h)$
                                        & $K_0 l_k (\nabla ^2 h) \phi _{t0}$ is the curvature-dependent \\
                 &                      & conformational current\\
$\gamma _a$      & $\gamma + n_0p_AK_0\phi _{t0}$
                                        & surface tension renormalized by active force dipoles \\
$\kappa _a $     & $\kappa _{eff} - n_0Q_AK_0l_K\phi _{t0}$
                                        & bending rigidity renormalized by curvature-induced \\
                 &                      & active quadrupoles \\
$c_a$            & $(\kappa _{eff} c_{eff}^h - n_0 Q_AK_0)/\kappa _{eff}$ %
                                        & inclusions-curvature coupling renormalized by \\
                 &                      & curvature-independent active force quadrupoles \\
\end{tabular}
\end{ruledtabular}
\end{table}

\newpage

%
% Table 3: typical magnitudes of parameters related to active forces
%
\begin{table}[h]
\caption{typical magnitudes of parameters that are important for
stability analysis \label{table:active}}
\begin{ruledtabular}
\begin{tabular}{c|l|l}
symbol         &  physical meaning          & typical magnitude\\ \hline %
$p_A $         &  active force dipole       & $|p_A|/a \lesssim {\rm pN \cdot ms}$ \\
$Q_A $         &  active force quadrupole   & $|Q_A|/a^2 \lesssim {\rm pN \cdot ms}$ \\
$K_0 $         &  characteristic time scale for curvature-independent & $|K_0| \lesssim {\rm ms}$\\
               &  inclusion conformational-change cycle &  \\
$K_0 l_K \nabla ^2 h $ &  characteristic time scale of curvature-induced & $|K_0 l_K \nabla ^2 h| \lesssim {\rm ms}$\\
               &  inclusion conformational-change cycle
                                            &  \\
$\phi _{t0}$   &  average areal fraction of the inclusions
                                            & $\phi _{t0} \lesssim 0.1 $ \\
$\chi _{eff}$  &  effective susceptibility of the inclusions
                                            & $\chi _{eff} \sim k_BT/{\rm nm^2}$ \\
$\kappa _{eff}$&  effective membrane curvature elastic modulus
                                            & $\kappa _{eff} \lesssim 10 k_BT$ \\
$c_{eff}^h, \ c_{eff}^\phi$
               &  effective inclusion-curvature elastic coupling
               constants
                                            & $|c_{eff}^h|, \
                                            |c_{eff}^{\phi}| \lesssim
                                            0.1 nm^{-1}$ \\
\end{tabular}
\end{ruledtabular}
\end{table}

\newpage

{\large \bf Figure Captions}
\begin{itemize}
\item Figure 1. Force
centers of the active force dipoles for $\beta \rightarrow \alpha$
transitions are distributed on surfaces described by $h+z_{\alpha
\beta}^{(u)}$, and $h-z_{\alpha \beta}^{(d)}$.  These surfaces and
$h({\bf r})$ are parallel surfaces, the areas of parallel surface
elements from top to bottom are $(1-z_{\alpha \beta}^{(u)} \nabla ^2
h) dA$, $dA$, and $(1+z_{\alpha \beta}^{(d)} \nabla ^2 h)dA$.

\item Figure 2. Schematics of the
conformational transitions of an inclusion with five internal
states.  When one considers the dynamics of the system at lengths
large compared to $\mathcal{O}(10 \ {\rm nm})$, the strength of the
conformational current of an inclusion with characteristic time
scale comparable to typical ion pumps is described $K_s \phi _t$.

\item Figure 3. Phase diagrams for membranes containing active inclusions.
$\gamma = 10^{-4}k_BT/{\rm nm}^2$, $k_2 = 250 k_BT \cdot {\rm
nm}^2$, $\phi _{t0}=0.1$, $M_{eff} \chi_{eff} = 1 \mu {\rm m}^2/s$,
$\eta = 10^{-3} {\rm kg/m \cdot s}$. Long-dashed curve indicates the
onset of wave instability, short-dashed curve indicate the onset of
Turing instability, solid line indicates the onset of long
wavelength instability.  The phase boundaries are determined by
taking $\kappa _{eff} = 5k_BT$ and $c_{eff}^{h}=c_{eff}^{\phi} = 0.1
{\rm nm}^{-1}$ at the onset of all instabilities. (a) $p_A/a=1 {\rm
pN \cdot ms} $, $Q_A/a^2=1 {\rm pN \cdot ms}$. For inclusions with
$l_K/a$ greater than the slope of the dashed line, the system has
Turing-type instability at large energy input, otherwise the system
has wave instability. (b) $p_A/a=1 {\rm pN  \cdot ms} $, $Q_A/a^2=-1
{\rm pN \cdot ms}$. In this case, only Turing-type instability
occurs at high energy input. (c) $p_A/a=- 1 {\rm pN \cdot ms} $,
$Q_A/a^2=1 {\rm pN \cdot ms}$. For inclusions with $l_K/a$ greater
than the slope of the dashed line, the system has Turing-type
instability at large energy input, otherwise the system has long
wavelength instability. (d) $p_A/a=-1 {\rm pN \cdot ms} $,
$Q_A/a^2=-1 {\rm pN \cdot ms}$. For inclusions with $|l_K/a|$
greater than the absolute value of the slope of the dashed line, the
system has Turing-type instability at large energy input, otherwise
the system has long wavelength instability.

\item Figure 4.  Mechanism for
Turing-type instability: (i) inclusions are attracted to regions
with preferred curvature, (ii) curvature-dependent active force
quadrupoles generate membrane curvature that inclusions prefer. This
positive feedback leads to a Turing-type instability.

\item Figure 5. characteristic wavelengths at the
onset of instabilities for $\gamma = 10^{-4}k_BT/{\rm nm}^2$, $k_2 =
250 k_BT \cdot {\rm nm}^2$, $\phi _{t0}=0.1$, $M_{eff} \chi_{eff} =
1 \mu {\rm m}^2/s$, $\eta = 10^{-3} {\rm kg/m \cdot s}$.  $\kappa
_{eff} = 5k_BT$, and $c_{eff}^{h}=c_{eff}^{\phi} = 0.1 {\rm
nm}^{-1}$.  The characteristic wavelengths are on the order of $100
{\rm nm}$ for all cases except at the vicinity of long-wavelength
instabilities, where the characteristic lengths increase abruptly.
(a) thin curves: $p_A/a=1 {\rm pN \cdot ms} $, $Q_A/a^2=1 {\rm pN
\cdot ms}$ (dashed curve: wave instability, solid curve: Turing-type
instability); thick curves: $p_A/a=0.1 {\rm pN \cdot ms} $,
$Q_A/a^2=1 {\rm pN \cdot ms}$ (dashed curve: wave instability, solid
curve: Turing-type instability). (b) dashed curve: $p_A/a=-1{\rm pN
 \cdot ms}$, $Q_A/a^2=1 {\rm pN \cdot ms}$; and solid
curve: $p_A/a=-1{\rm pN \cdot ms}$, $Q_A/a^2=-1 {\rm pN \cdot ms}$.

\item Figure 6. Mechanism for
wave instability: (i) inclusions are attracted to regions with their
preferred curvature (ii) in inclusion-rich domains, active force
quadrupoles produce membrane curvature that inclusions dislike, the
inclusions leave inclusion-rich domains and the membrane
configuration goes back to (i).

\item Figure 7. Characteristic time
scale at the onset of wave instability for $\gamma =
10^{-4}k_BT/{\rm nm}^2$, $k_2 = 250 k_BT \cdot {\rm nm}^2$, $\phi
_{t0}=0.1$, $M_{eff} \chi_{eff} = 1 \mu {\rm m}^2/s$, $\eta =
10^{-3} {\rm kg/m \cdot s}$.   $\kappa _{eff} = 5k_BT$, and
$c_{eff}^{h}=c_{eff}^{\phi} = 0.1 {\rm nm}^{-1}$.  Dashed curve:
$p_A/a=1 {\rm pN \cdot ms} $, $Q_A/a^2=1 {\rm pN \cdot ms}$; solid
curve: $p_A/a=0.1 {\rm pN \cdot ms}$, $Q_A/a^2=1 {\rm pN \cdot ms}$.
The magnitude of $l_K/a$ for the solid curve is indicated at the top
of the frame, and the magnitude of $l_K/a$ for the dashed curve is
indicated at the bottom.

\item Figure 8. A conical inclusion in state $\alpha$, $l_{\alpha}$ is the thickness
of the inclusion, $r_{\alpha}^{out}$ is the outer radius,
$r_{\alpha}^{in}$ is the inner radius of the inclusion.

\item Figure 9. Regions rich in
positive $c_{\alpha}$ inclusions tend to have $\nabla ^2 h<0$.
\end{itemize}

\newpage

%
% Here are the figures
%

%
% Figure 1
%

\begin{figure}[h]
%%\begin{left}
%\rotatebox{270}{
\epsfxsize= 4.5 in \epsfbox{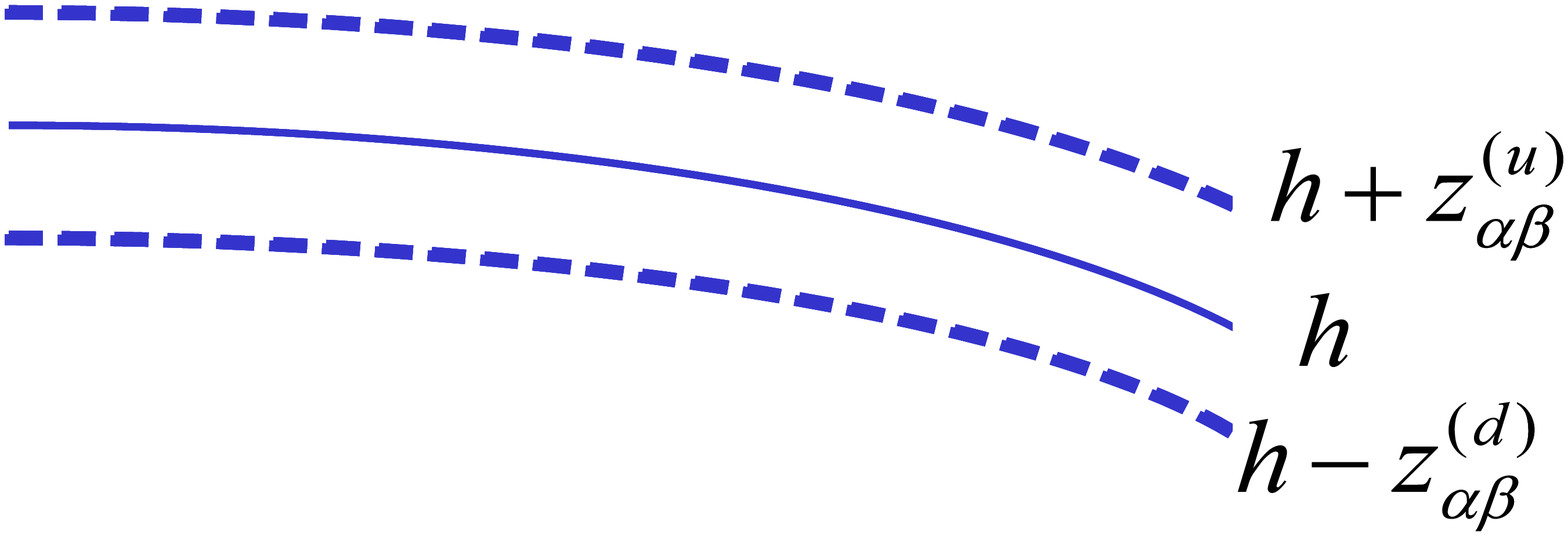} \caption{}
\label{fig:dipole_density}
\end{figure}

%
% Figure 2
%

\begin{figure}[h]
%\begin{left}
%\rotatebox{270}{
\epsfxsize= 4.5 in \epsfbox{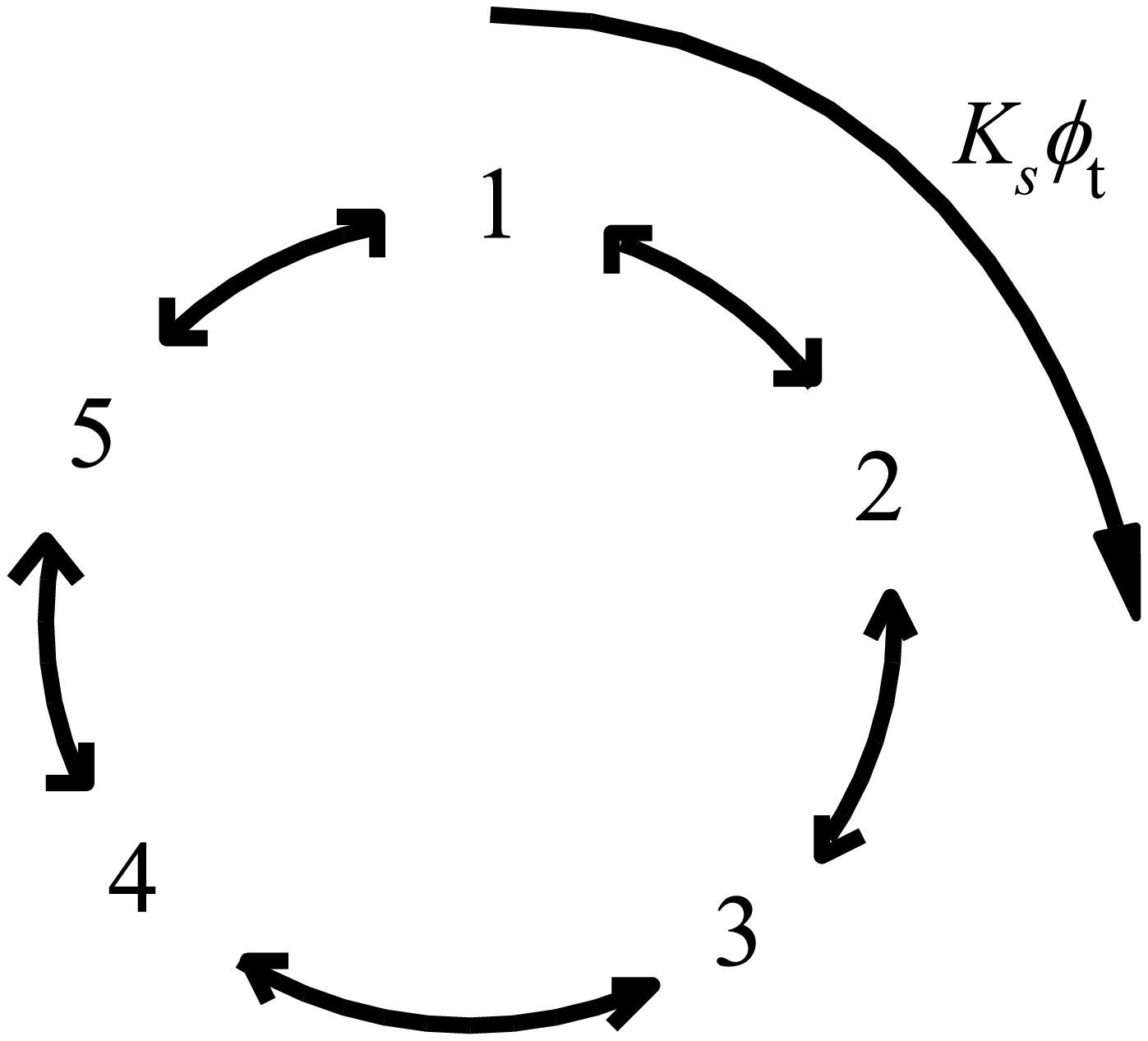} \caption{}
\label{fig:states}
\end{figure}

%
% Figure 3
%

\begin{figure}[h]
%%\begin{left}
%\rotatebox{270}{
\epsfxsize= 3 in \epsfbox{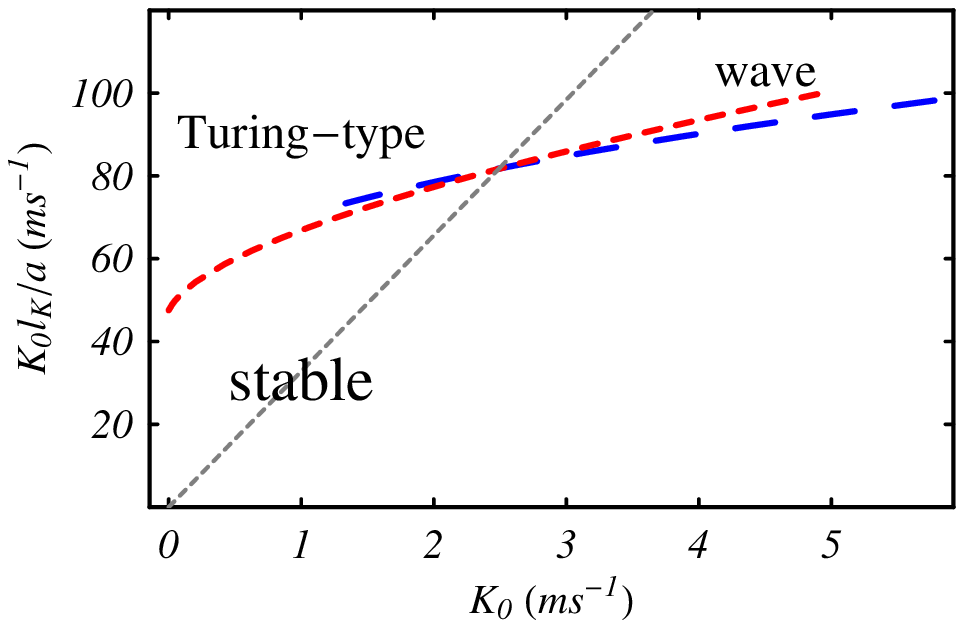}~(a) %
\epsfxsize= 3 in \epsfbox{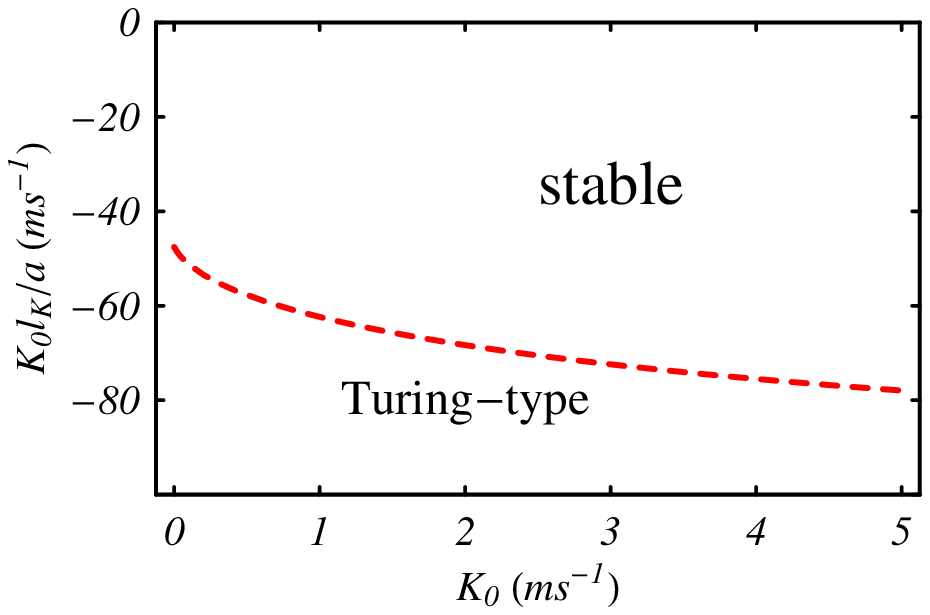}~(b) %
\epsfxsize= 3 in \epsfbox{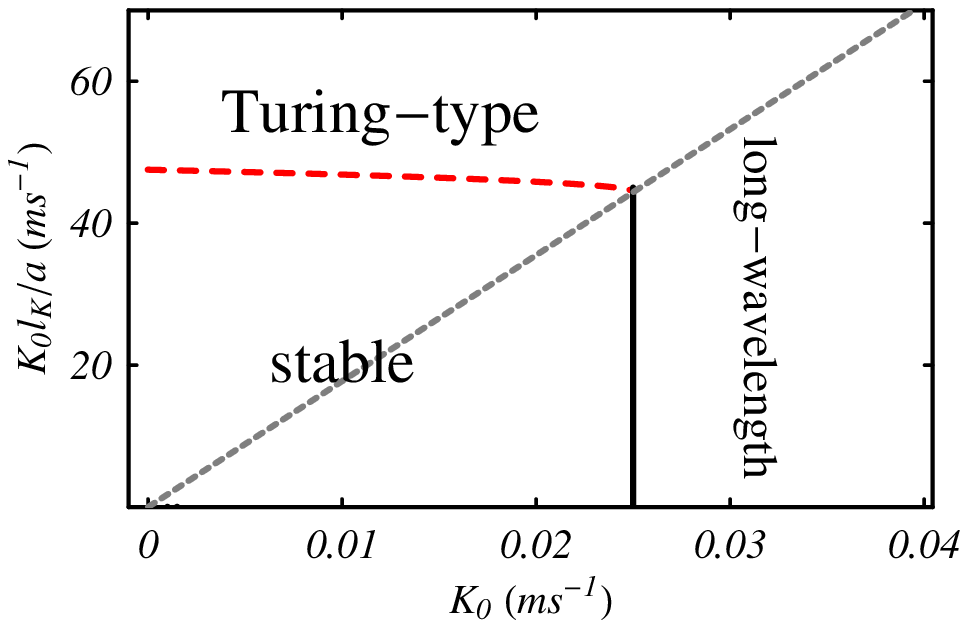}~(c) %
\epsfxsize= 3 in \epsfbox{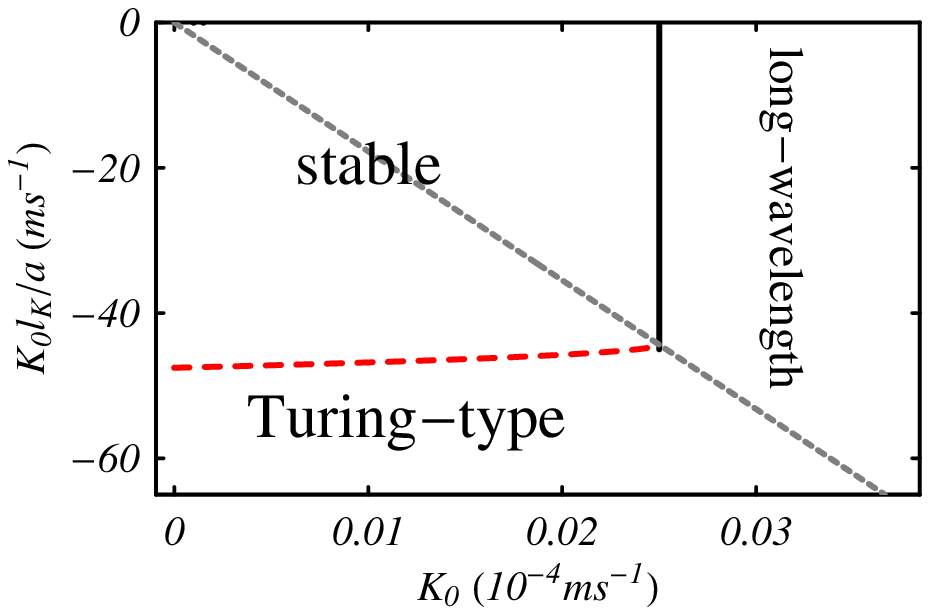}~(d) %
\caption{} \label{fig:phase}
\end{figure}

%
% Figure 4
%

\begin{figure}[h]
%%\begin{left}
%\rotatebox{270}{
\epsfxsize= 4.5 in \epsfbox{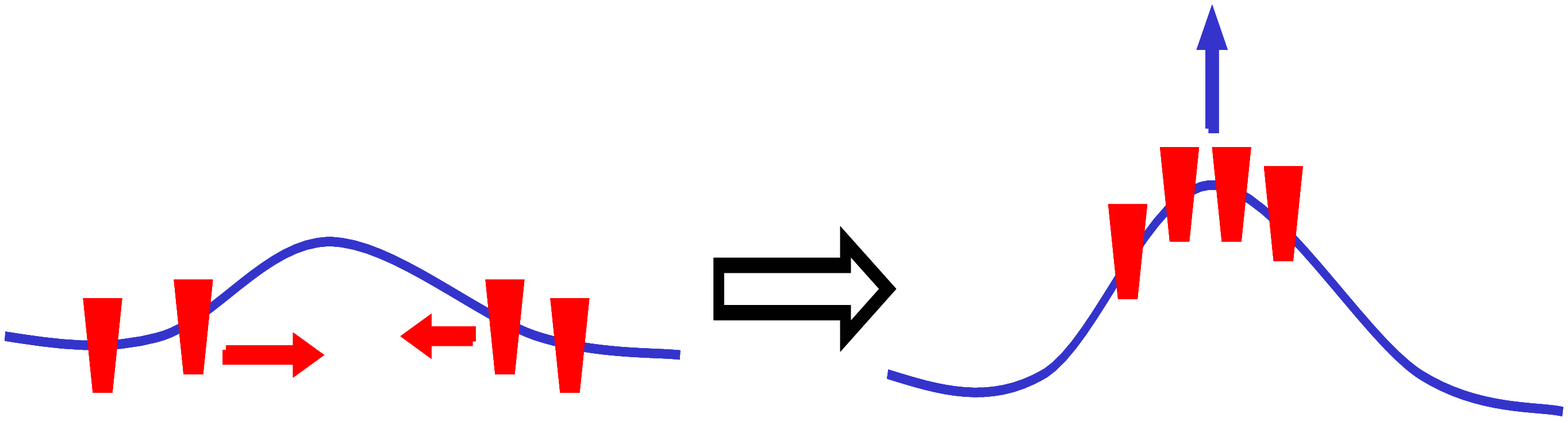} \caption{}
\label{fig:static_inst}
\end{figure}

%
% Figure 5
%

\begin{figure}[h]
%%\begin{left}
%\rotatebox{270}{
\epsfxsize= 4.5 in \epsfbox{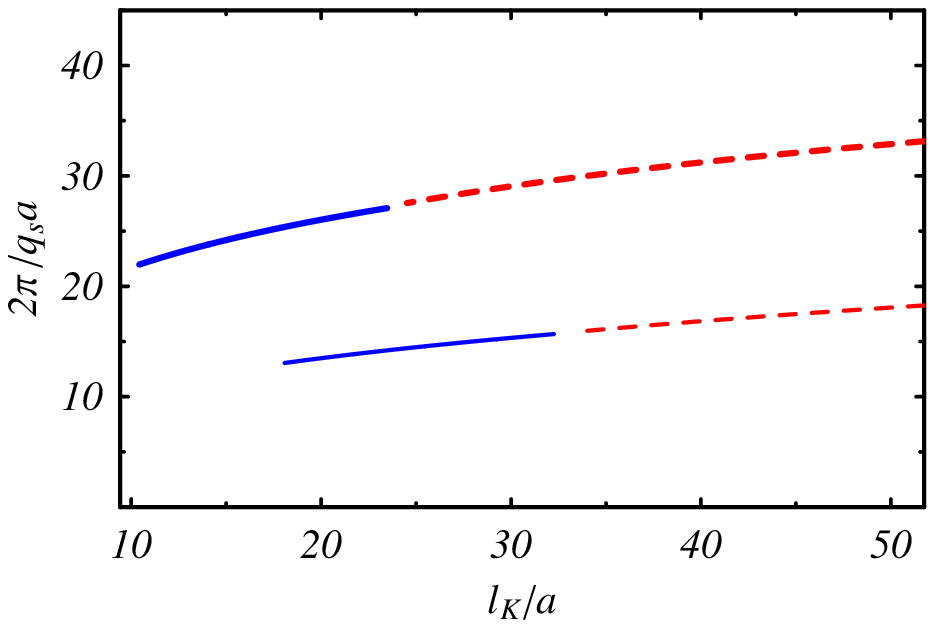} \epsfxsize=4.5in (a)
\vskip 1cm %
\epsfbox{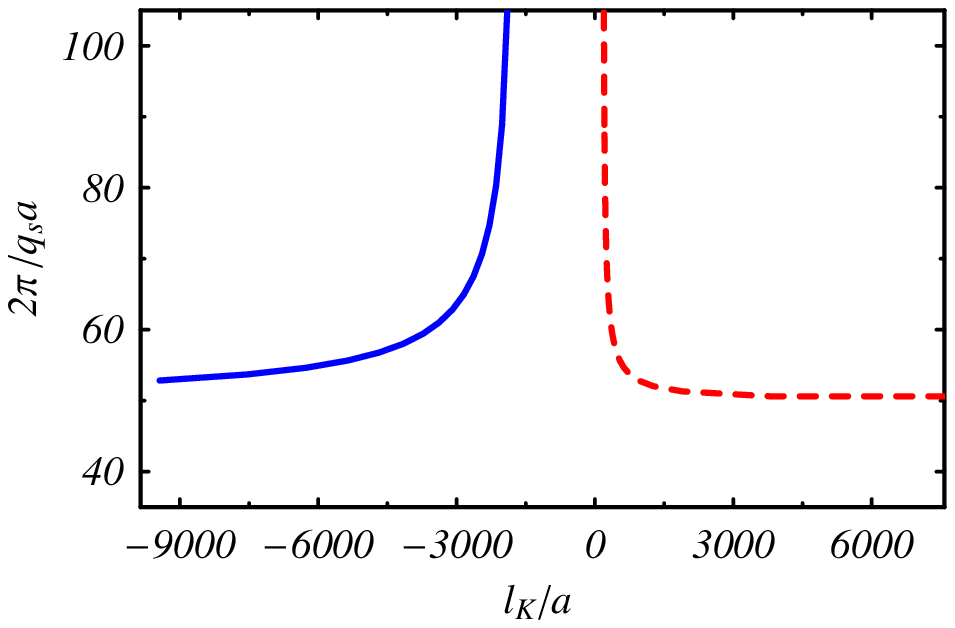} (b) \vskip 1cm %
\caption{}\label{fig:length}
\end{figure}

%
% Figure 6
%

\begin{figure}[h]
%%\begin{left}
%\rotatebox{270}{
\epsfxsize= 4.5 in \epsfbox{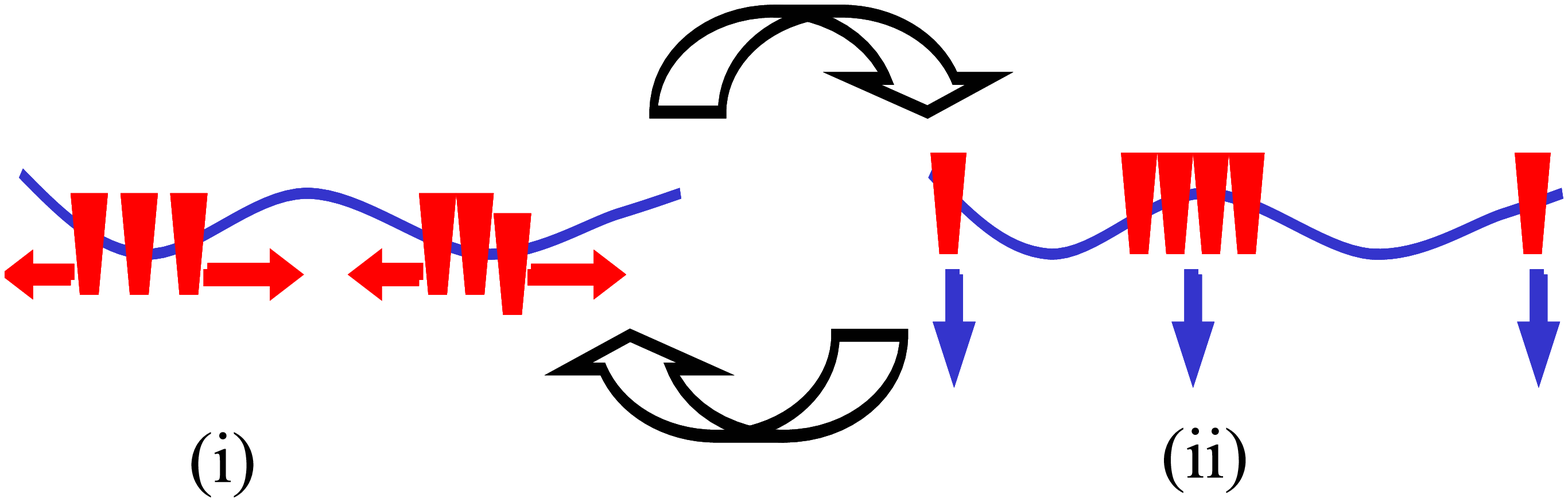} \caption{}
\label{fig:wave_inst}
\end{figure}

%
% Figure 7
%

\begin{figure}[h]
%\begin{left}
%\rotatebox{270}{
\epsfxsize= 4.5 in \epsfbox{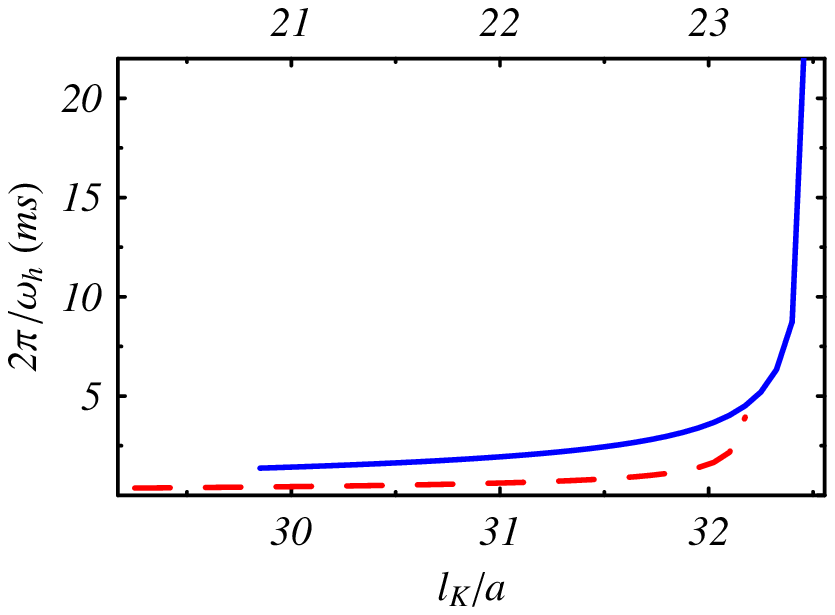} \caption{} \label{fig:time}
\end{figure}

%
% Figure 8
%

\begin{figure}[h]
%\begin{left}
%\rotatebox{270}{
\epsfxsize= 4.5 in \epsfbox{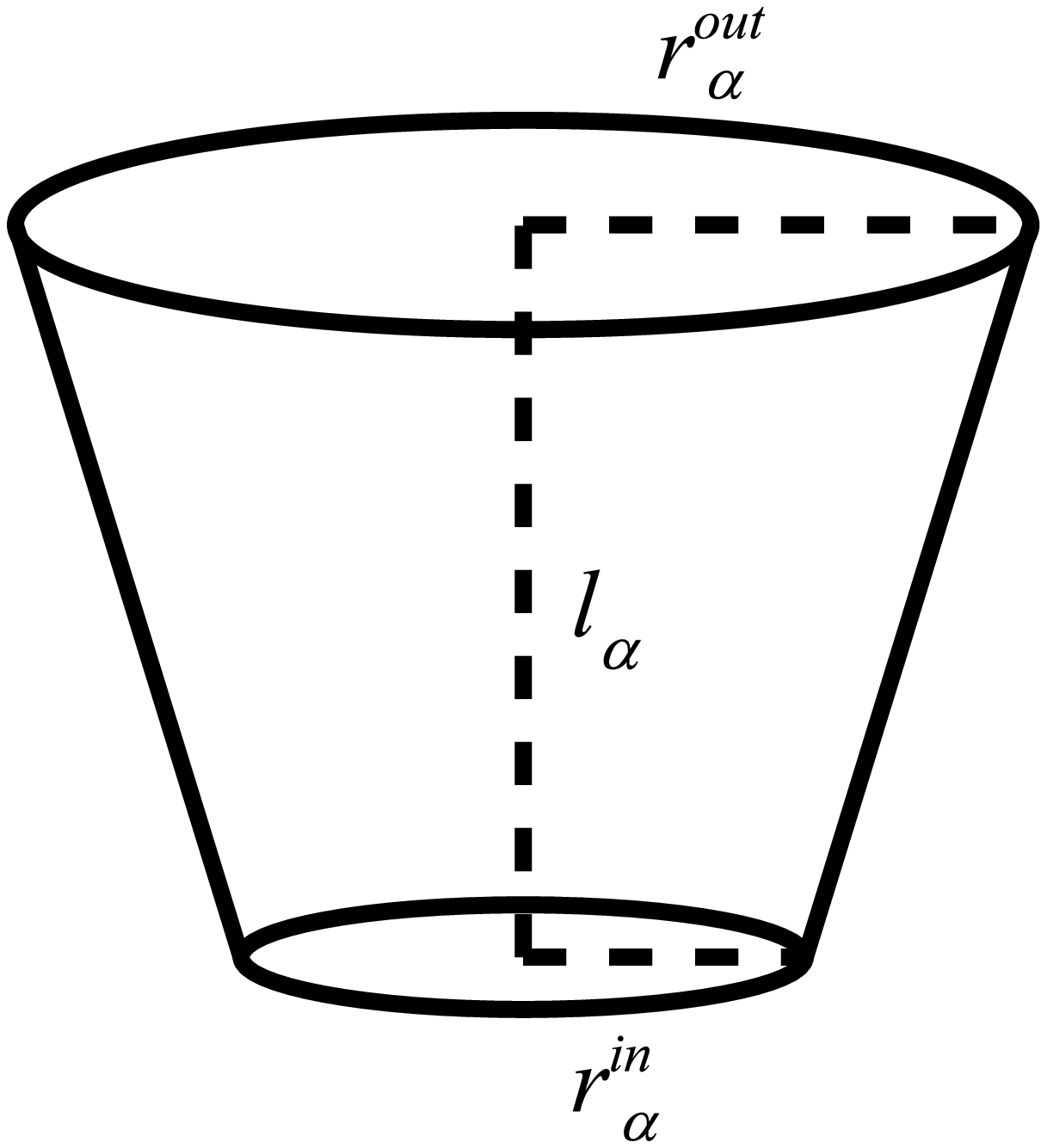} \caption{}
\label{fig:c_alpha}
\end{figure}

%
% Figure 9
%
\begin{figure}[h]
%\begin{left}
%\rotatebox{270}{
\epsfxsize= 2.5 in \epsfbox{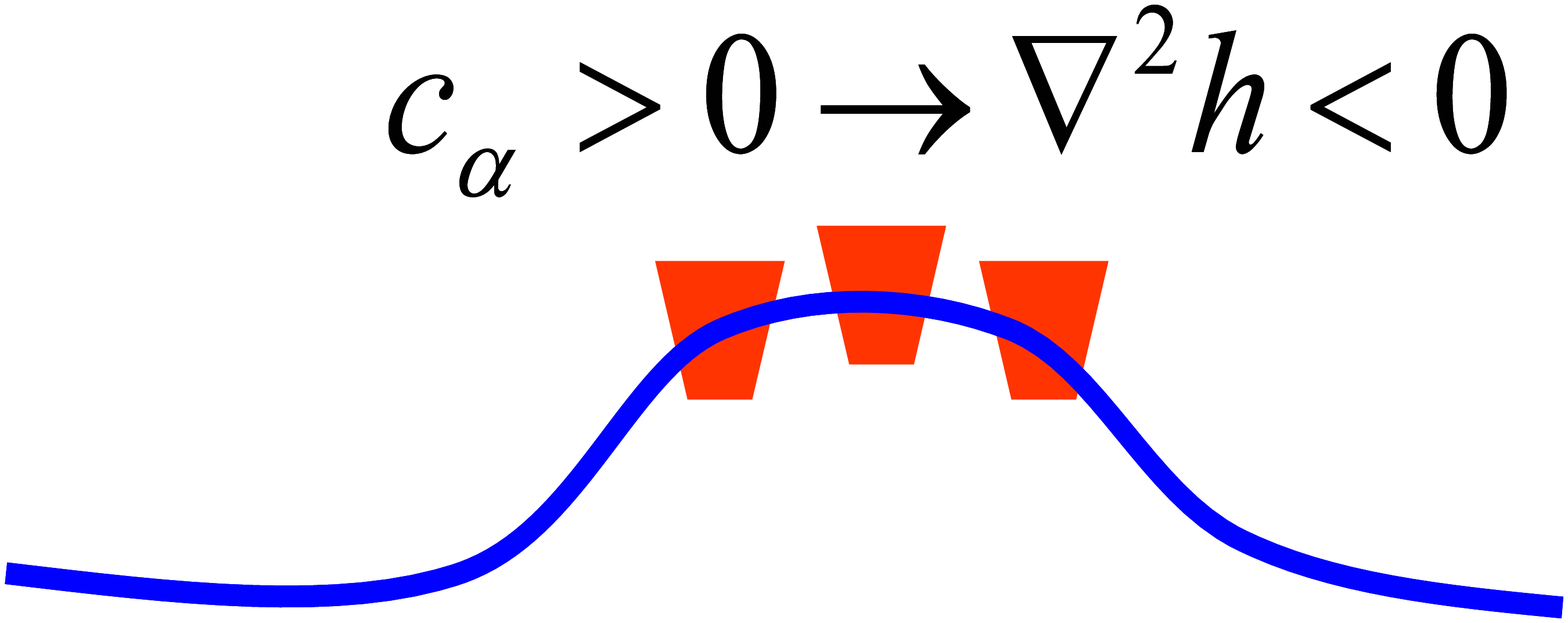} \caption{}
\label{fig:e_couplings}
\end{figure}


\newpage

\begin{references}

\bibitem{ref:Lodish_book} H. Lodish {\it et al.}, {\it Molecular
cell biology}, 3rd ed. (W.H. Freeman, New York, 1995), 3rd ed.

\bibitem{ref:06_JP_Sackmann} E. Sackmann, J. Phys.: Condens. Matter,
{\bf 18}, R785 (2006).

\bibitem{ref:97_Adv_Phys_Seifert} U. Seifert, Adv. Phys, {\bf 46}, 13
(1997).

\bibitem{ref:96_EPL_Prost} J. Prost and R. Bruinsma, Europhys.
Lett., {\bf 33}, 321 (1996).

\bibitem{ref:01_PRE_Prost} J.-B. Manneville, P. Bassereau, S.
Ramaswamy, and J. Prost, Phys. Rev. E, {\bf 64}, 021908 (2001).

\bibitem{ref:05_PRL_Prost} P. Girard, J. Prost, and P. Bassereau,
Phys. Rev. Lett., {\bf 94}, 088102 (2005).

\bibitem{ref:08_Prost} M.M. El Alaoui Faris, D. Lacoste, J.
P\`{e}cr\`{e}aux, J-F. Joanny, J. Prost, and P. Bassereau, Phys.
Rev. Lett., {\bf 102}, 038102 (2009).

\bibitem{ref:00_PRL_Ramaswamy}  S. Ramaswamy, J. Toner, and J. Prost, Phys.
Rev. Lett., {\bf 84}, 3494 (2000).

\bibitem{ref:04_PRL_Chen} H-Y. Chen, Phys. Rev. Letts., {\bf 92},
168101, (2004).

\bibitem{ref:06_PRE_Chen} C-H. Chen and H-Y. Chen, Phys. Rev. E,
{\bf 74}, 051917 (2006).

\bibitem{ref:92_BPJ_Needham} Needham. D. and R. M. Hochmuth, Biophys. J. {\bf
61}, 1664, (1992).

\bibitem{ref:95_BPJ_Strey} H. Strey, M. Peterson, and E. Sackmann, Biophys. J. {\bf
69}, 478, (1995).

\bibitem{ref:equilibrium} See, for example, R.R. Netz and P. Pincus,
Phys. Rev. E, {\bf 52}, 4114 (1995).

\bibitem{ref:04_BPJ_Turner_Sens} P. Sens and M.S. Turner, Biophys.
J., {\bf 86}, 2049 (2004).

\bibitem{ref:Lomholt} M.A. Lomholt, Phys. Rev. E {\bf 73}, 061913
(2006).

\bibitem{future} H-Y. Chen and A.S. Mikhailov, unpublished.

\end{references}
\end{document}